\documentclass[twocolumn,showpacs,amsmath,amssymb,a4paper,aps]
{revtex4-1}

\def\@dotsep{4.5}
\makeatother

\usepackage{bm}
\usepackage{graphicx}

\usepackage{amssymb}
\usepackage{amsmath}
 
\newcommand{\BEA}{\begin{eqnarray}}
\newcommand{\BEAN}{\begin{eqnarray*}} 
\newcommand{\EEA}{\end{eqnarray}}
\newcommand{\EEAN}{\end{eqnarray*}}

\begin{document}


\title{Global {\em ab initio} potential energy surface for the 
O$_2(^3\Sigma^-_g)$ + N$_2(^1\Sigma^+_g)$ interaction. Applications
to the collisional, spectroscopic, and thermodynamic properties of
the complex}

\author{Massimiliano Bartolomei\footnote{Electronic mail: 
maxbart@iff.csic.es}, Estela Carmona-Novillo, Marta I. Hern\'{a}ndez, 
Jos\'e Campos-Mart\'{\i}nez\footnote{Electronic mail: 
j.campos.martinez@csic.es} }
\affiliation{Instituto de F\'{\i}sica Fundamental, (IFF-CSIC)
Consejo Superior de Investigaciones Cient\'{\i}ficas, Serrano 123, 
28006 Madrid, Spain}
\author{Robert Moszy{\'n}ski }
\affiliation{Quantum Chemistry Laboratory,
Faculty of Chemistry,
University of Warsaw,
L. Pasteura 1 02-093 Warszawa, Poland}


\begin{abstract}
A detailed characterization of the interaction between the most abundant
molecules in air is important for the understanding of a variety of phenomena
in atmospherical science.  
A completely {\em ab initio} global potential energy surface (PES) for the
O$_2(^3\Sigma^-_g)$ + 
N$_2(^1\Sigma^+_g)$ interaction is reported for the first time. It has been obtained
with the symmetry-adapted perturbation theory utilizing a density functional
 description of monomers [SAPT(DFT)]  extended to treat the interaction
 involving  high-spin open-shell complexes. 
 The computed interaction energies of the complex are in a good agreement with those
 obtained by using the 
spin-restricted coupled cluster methodology with singles, doubles and noniterative
triple excitations  [RCCSD(T)]. 
 A  spherical harmonics expansion containing a large
number of terms due to the anisotropy of the interaction has been built
from the {\em ab initio} data. The radial coefficients of the expansion are
matched in the long range with the analytical functions based on the recent 
{\em   ab initio} calculations of the electric properties of the monomers 
[M. Bartolomei et al., J. Comp. Chem., {\bf 32}, 279 (2011)]. 
 The PES is tested against the second virial coefficient $B(T)$ data and the integral
 cross sections measured 
with rotationally hot effusive beams, leading in both cases to a very good agreement.
The first bound states of the complex have been computed and relevant
spectroscopic features of the interacting complex are reported.
A comparison with a previous experimentally derived PES is also provided.
\end{abstract}


\maketitle

\section{ Introduction}
\label{intro}

Simple molecular gases interact in our atmosphere and the outer space
through weak intermolecular forces giving rise to many different processes 
that roughly speaking are characterized by the energy transfer and 
momentaneous rearrangements of their isolated charge distributions.  
It is, therefore, of a paramount interest an accurate description of 
these weak interactions and properties that appear as a consequence.  
Among these simple molecular species, O$_2$ and N$_2$ are central 
because of their abundance in our atmosphere, and their importance for life.

Another important consequence of the weak intermolecular forces is the
existence of weakly bound complexes, mainly dimers.  In some cases as, 
in the  O$_2$ dimer, its existence was already suggested as early as 1924 
when Lewis\cite{Lewis:24} proposed the existence of a strong bound between two
oxygen molecules.  In search of such a dimer Welsh
et. al\cite{Welsh-dimer:49} found much stronger and universal bands, which
are known as Collision-Induced Absorption\cite{cia-frommhold} (CIA).  Both the
dimeric complex and CIA are close and related phenomena that are important not
only for spectroscopic purposes\cite{frommhold-13-cia-astro}, 
but for the energy transfer rates and other
observables that are key ingredients in the atmospheric 
modeling\cite{klemperer-06-molclus-atmos} and 
many other areas in which the delicate balance between radiative and 
non-radiative processes is neccesary to reach an acceptable agreement with
observation. 

Since the importance of the previously mentioned effects are intimately 
related to the gas density, the knowledge of an accurate potential energy 
surface (PES) that could describe short and long-range behavior at once 
becomes evident.  

For the oxygen dimer, we have recently succeeded in producing a
global potential energy surface with {\em ab initio} calculations at a  
high level of
theory.  In particular, for the quintet state multiplicity a 
restricted coupled-cluster theory with singlet, doubles, and perturbative
triple excitations [RCCSD(T)] was successfully used\cite{max-jcp:08} 
to compute this PES.   This state is well represented by a single
configuration while the singlet and triplet states are inherently
multiconfigurational and therefore the CCSD(T) cannot be applied.
A scheme was proposed\cite{maxpccp:08} to sort out this
problem that was based on a difference procedure involving multireference
electronic structure methods.   The result is a global potential
energy surface\cite{O2O2pes:10} that yields excellent results as compared 
with all experimental data available and both for the short and 
long range\cite{max-lr:10}.  In the case of the
nitrogen dimer, although it is not an open-shell system it was not easy 
to accomplish an accurate description until recently.  
Currently, there are several {\em ab initio} PES\cite{Bussery:07} that have been computed 
using Symmetry Adapted Perturbation Theory (SAPT)\cite{sapt:94}, and 
that have been very recently used\cite{hart-buss-cia-n2:13} to study 
CIA for nitrogen. 

The data for the heterodimer $N_2-O_2$ are scarcer, even though
an  accurate description of the interactions is of primordial
relevance to interpret data from observation and remote sensing and
atmospheric modeling.  One of the
few available PES is an experimentally derived potential energy 
surface\cite{Aquilanti:03} (Perugia-PES, from now on), which was obtained from
molecular beam experiments and gives an excellent answer when compared
with available experimental data.

 Here, we employ SAPT(DFT) to calculate for the first time an accurate 
 {\em ab initio} PES that is further expanded in order to have a functional 
 form ready to use in dynamical calculations.  The performance of the newly
 computed PES is compared with previous ones as well as tested against
 experimental data. It should be stressed that the SAPT(DFT) approach is 
particularly well suited for the calculations of global potential energy 
surfaces for weakly bound complexes. As shown in Ref. \cite{SAPTDFT:08} 
the accuracy of the SAPT(DFT) results is often comparable to RCCSD(T), 
while the computational cost is much lower.


The paper is organized as follows. Section \ref{pes} gives the details of 
the PES  calculations, including the angular expansion and the long-range behavior.
In Section \ref{results} we discuss the topography of the potential energy
surface and compare the computed second virial coefficients and integral
cross sections with the available experimental data. In Section \ref{bound}
we report calculations of the bound rovibrational levels. Finally, in Section 
\ref{conclu} we conclude our paper.

\section{ The O$_2$-N$_2$  PES: calculations}
\label{pes}

\begin{figure}[t]
\includegraphics[width=8.5cm,angle=0.]{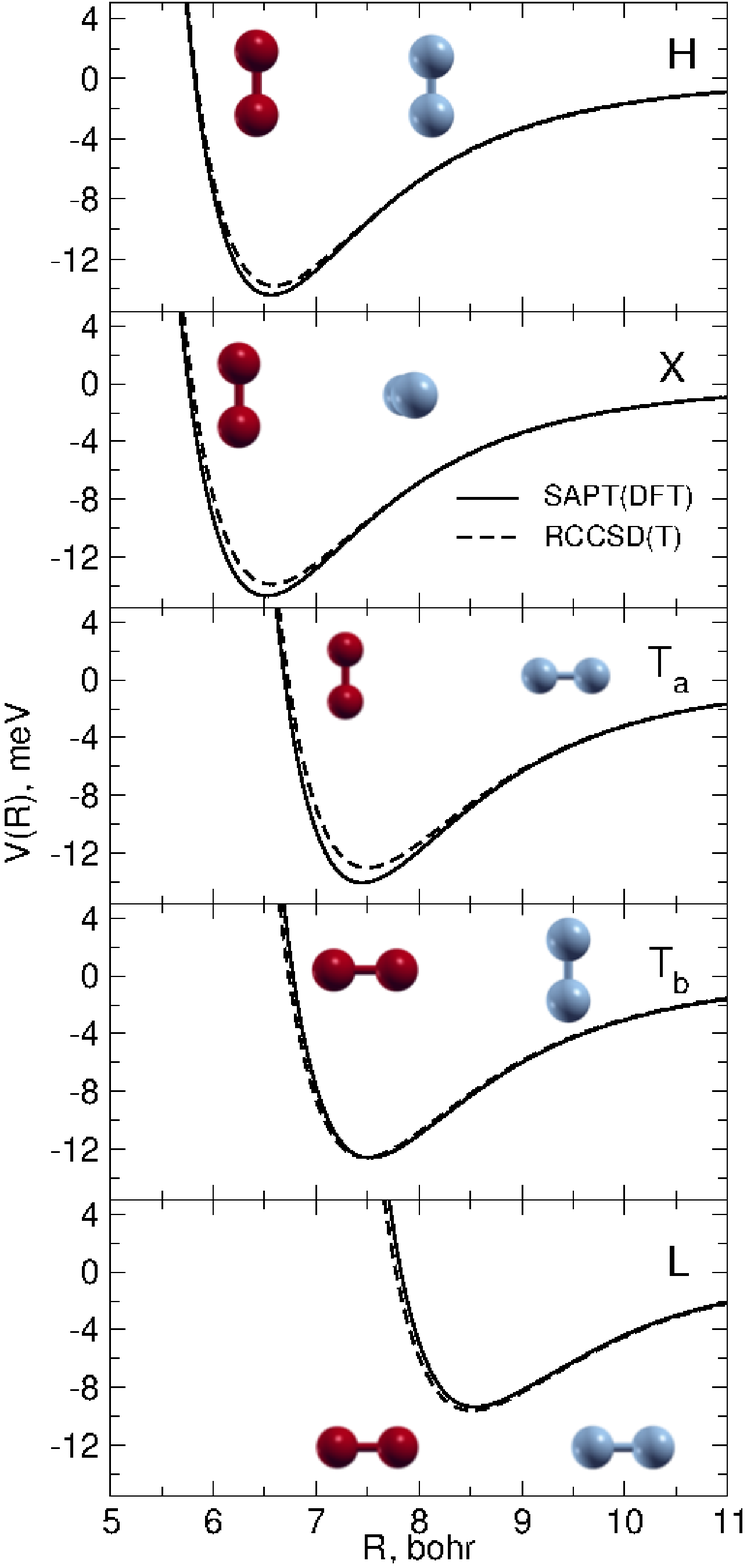}
\caption[]{Comparison between  SAPT(DFT) and RCCSD(T) intermolecular
  potentials at five limiting orientations of O$_2$-N$_2$ as obtained by using
  the aug-cc-pVTZ plus bond function  basis set (see text). The different 
  arrangements are depicted in the insets, where O and N atoms are symbolized 
 by red and light blue balls, respectively.}
\label{fig1}
\end{figure}

Diatom-diatom Jacobi vectors, ${\bf r}_a$, ${\bf  r}_b$ and ${\bf
  R}$, are used, where  ${\bf r}_a$ and ${\bf  r}_b$ are the interatomic
vectors of O$_2$ and N$_2$, respectively, and ${\bf R}$ is the vector joining
the centers of mass of the diatoms. Thus, the six internal coordinates
determining the O$_2$-N$_2$ PES are the
distances $R$, $r_a$, and $r_b$, and $\theta_a$, $\theta_b$, and $\phi$, the
angles formed by ${\bf R}$ and ${\bf r}_a$, ${\bf R}$  and ${\bf r}_b$,  and
the torsional angle, respectively. In the PES reported here, the intramolecular
distances have been fixed at the O$_2$ and N$_2$ equilibrium distances, $r_a$=
2.28 and $r_b$= 2.08 bohr, respectively, so that we describe the 
interaction between the diatoms in their ground vibrational states. 

As done previously\cite{max-jcp:08,O2O2pes:10} we have  considered the
geometries generated from 9 Gauss-Legendre quadrature 
points in the $-1 \leq cos\theta_a,cos\theta_b  \leq 1$ range and from 5 
Gauss-Chebyshev points for $0 \leq \phi  \leq \pi$. 
An initial range in the torsional angle from [0:$2\pi$] has been reduced to 
[0:$\pi$] by use of the permutation-inversion symmetry of this
system\cite{vdABrocks:87}, which also allows us to reduce the initial 
9$\times$9$\times$5 grid to a 5$\times$5$\times$5 grid 
($-1 \leq cos\theta_a,cos\theta_b  \leq 0$). If we further consider the 
following equivalences due to symmetry
$(\theta_a,\pi/2,\phi) \equiv (\theta_a,\pi/2,\pi-\phi)$ and 
$(\pi/2,\theta_b,\phi) \equiv (\pi/2,\theta_b,\pi-\phi)$, 
we are finally left 
with 107 ``irreducible'' geometries. For each angular arrangement, 18 points 
in the intermolecular coordinate $R$, ranging from 16.0 to 5.0 bohr, were 
considered.

\subsection{ SAPT(DFT)  {\em ab initio} calculations}  

\begin{table}
\caption{Energies (in meV) for five limiting configurations at
  intermolecular distances close to the minima (in parenthesis, in bohr),
as obtained from SAPT(DFT) with aug-cc-pVTZ and aug-cc-pVQZ  basis sets plus 
bond functions (the same in both cases, see text). 
}
\label{table1}
\begin{tabular}{lccccc}
& \; {\sf H}  (6.5) & \; {{\sf X}  (6.5)} & \; {\sf T$_a$}  (7.5) & \; {\sf
    T$_b$}  (7.5) & \; {\sf L}  (8.5) \\
\hline
VTZ\,\,\, &  14.344  & 14.685 & 14.004 & 12.580 & 9.324\\
VQZ\,\,\, &  14.549  & 14.909 & 14.104 & 12.723 & 9.322\\
\end{tabular}
\end{table}

Since the O$_2$ and N$_2$ monomers have electronic spins $S_a = 1$ and  $S_b= 0$, 
respectively, the intermolecular potential does not depend on the spin
couplings between the monomers and therefore accurate single reference approaches, such as 
SAPT(DFT) and spin-restricted  RCCSD(T), can be applied. 
The SAPT(DFT) calculations were performed using the methodology of
Ref. \cite{SAPTDFT:08} which is suitable to obtain the interactions in
high-spin dimers formed by open-shell monomers. The aug-cc-pVTZ and aug-cc-pVQZ
basis sets\cite{Dunning}, both supplemented by the bond function set
[3s3p2d1f] developed 
by Tao\cite{tao:92} and placed in the middle of the intermolecular distance, have been used in
the calculations. The monomer DFT calculations employed the PBE0 hybrid
functional\cite{PBE0} and were done using the DALTON\cite{Dalton} program. We
applied the Fermi-Amaldi-Tozer-Handy asymptotic
correction\cite{Fermi:34,Handy:98} with the ionization potentials of  
 0.479 and 0.613 a.u.\cite{zuchowski:08} (for $\alpha$ and $\beta$ electrons,
 respectively) for O$_2$ and 0.5726 a.u.\cite{NIST} for N$_2$.
The SAPT(DFT) calculations used the open-shell SAPT(DFT) program of the
SAPT2008 code \cite{SAPT-code}. 

First, we have tested the basis set saturation and performances for 
 some limiting configurations of the interacting complex: linear ($D_{\infty h}$), 
T-shaped ($C_{2v}$), rectangular ($D_{2h}$) and crossed ($D_{2d}$)
arrangement, in the following, referred to as {\sf L}, {\sf T}, {\sf H} and
{\sf X}, respectively. We distinguish two T-shaped orientations: {\sf T$_a$},
where the O$_2$ intramolecular vector ${\bf r}_a$ is perpendicular to ${\bf
  R}$, and {\sf T$_b$}, where N$_2$ is the diatom oriented perpendicularly to
the intermolecular vector.  Results are reported in Table \ref{table1}. As can
be noticed both sets provide almost identical results: the largest
differences, corresponding to {\sf H} and {\sf X}  configurations, are within
1 \%. The analysis was extended to other intermolecular distances and it was
found that the discrepancies between both basis sets remain of the same order
of magnitude. We regard these deviations as acceptable, bearing in mind that
for a single geometry on a single processor  the aug-cc-pVTZ calculations are
three times faster than those employing the larger basis set. 

 In order to further check the performance of the SAPT(DFT) results, 
supermolecular spin-restricted RCCSD(T)/aug-cc-pVTZ (plus bond functions)
calculations have been performed for the same limiting configurations.
The counterpoise method\cite{Boys:70,Lenthe:87} was applied to correct for the basis 
set superposition error and the calculations were carried out with the 
{\sc Molpro}2006.1 package\cite{MOLPRO}. The comparison between the interaction
energies obtained from the two different levels of theory are shown in
Fig. \ref{fig1}. It can be noticed that SAPT(DFT) curves are somewhat more
attractive in the well region for the most attractive configurations ({\sf H},
{\sf X} and {\sf T$_a$}), while they are slightly more repulsive for the remaining
arrangements ({\sf T$_b$} and {\sf L}). Overall the agreement can be
termed as good since discrepancies are in general within 5 \%. It is 
noteworthy to stress the saving in computing time provided
by the SAPT(DFT) methodology. In fact we have estimated that for a single
point/single CPU, RCCSD(T) calculations would be about six times slower      
 than equivalent SAPT(DFT) ones if no symmetry is used (as it is for most of
 the geometrical arrangements). 

Supported by the above mentioned tests we decided to choose the
SAPT(DFT) methodology with the aug-cc-pVTZ basis supplemented with the
[3s3p2d1f] set of bond functions for all the 
geometries needed to obtain a complete and reliable representation of the
O$_2$-N$_2$ PES.

\subsection{Spherical harmonics expansion}

\begin{table}
\footnotesize
\renewcommand{\baselinestretch}{0.7}
\caption{Comparison of {\it ab initio} energies with those resulting from the
  spherical harmonic expansion of Eq. \ref{expansion} using the set of 39
  radial coefficients detailed in the text, for several orientations and
  intermolecular distances. 
}
\label{table2}
\renewcommand{\arraystretch}{0.5}
\begin{tabular}{lllrrrr}
\multicolumn{3}{c}{Geometry}   &    &    &    &   \\
 $\theta_a$  &  $\theta_b$ &  $\phi$   & $R$   & $V_{ab initio}$   &  $V_{expanded}$ & \%  \\
\multicolumn{3}{l}{(degrees)}   & (bohr)   &  (meV)  &   (meV) &   \\
\hline
   &   &     & 5.25   & 71.30   & 71.63  & 0.46   \\
\multicolumn{3}{c}{H} & 5.50   & 28.10   & 28.08 & 0.07   \\
 90  &  90 &  0   &  6.25  & -12.77     &  -12.84  & 0.55   \\
   &   &     &  6.50  & -14.34    &   -14.37  & 0.66  \\
   &   &     &  7.00  & -12.75    &   -12.75 & 0.21  \\
   &   &     & 12.00   & -0.513   &   -0.513  & 0.00  \\
   &   &     & 14.00   & -0.192   &   -0.192  & 0.00  \\
\hline
   &   &     & 5.25   & 57.20     & 55.25 & 3.41   \\
\multicolumn{3}{c}{X} & 5.50   & 20.64   &  20.02  & 3.00   \\
 90  &  90 & 90   &  6.25  & -13.59   &  -13.50   & 0.66  \\
   &   &     &  6.50  & -14.69    &   -14.63 & 0.41  \\
   &   &     &  7.00  & -12.80    & -12.78  & 0.16    \\
   &   &     & 12.00   & -0.546   & -0.546 & 0.00   \\
   &   &     & 14.00   & -0.209   & -0.209  & 0.00   \\
\hline
   &   &     &  5.25   & 577.57   & 577.84 & 0.05    \\
\multicolumn{3}{c}{T$_a$} & 5.50   & 341.08   & 340.70  & 0.11   \\
 90  &  0 &  0   &  7.25  & -13.52   &  -13.52 & 0.00  \\
   &   &     & 7.50   &  -14.00  & -14.01 & 0.07  \\
   &   &     & 8.00   &  -11.85   & -11.85 & 0.00  \\
   &   &     & 12.00   &  -0.954  & -0.954 & 0.00  \\
   &   &     & 14.00   &  -0.351  & -0.351 & 0.00    \\
\hline
   &   &     &  5.25   & 661.64   & 660.63 & 0.15 \\
\multicolumn{3}{c}{T$_b$} & 5.50   & 389.72   & 388.71  & 0.26 \\
 0  &  90 &  0   &  7.25  & -11.57   &  -11.56 & 0.09  \\
   &   &     & 7.50   &  -12.58  & -12.57 & 0.08  \\
   &   &     & 8.00   & -11.01   & -11.00 & 0.09  \\
   &   &     & 12.00   &  -0.928  & -0.928 & 0.00  \\
   &   &     & 14.00   &  -0.350  & -0.350 & 0.00    \\
\hline
   &   &     & 5.25   & 3332.44   & 3344.74 & 0.37 \\
\multicolumn{3}{c}{L} & 5.50   & 2381.61   & 2384.07  & 0.10   \\
 0  &  0 &  0   &  8.25  & -8.35   & -8.33 & 0.24   \\
   &   &     & 8.50   & -9.32   & -9.31  & 0.11   \\
   &   &     & 9.00   & -8.20   & -8.21  & 0.12   \\
   &   &     & 12.00   &  -1.129  & -1.128 & 0.09  \\
   &   &     & 14.00   &  -0.361  & -0.361 & 0.00  \\
\hline
   &   &     & 5.25   & 626.06   & 626.54 & 0.08  \\
\multicolumn{3}{c}{S$_{45}$} & 5.50   & 386.29 & 386.37 & 0.02 \\
 45  &  45   &  0   & 7.25   &  -9.40  & -9.38 & 0.21   \\
   &   &     &  7.50  & -11.40   & -11.40 & 0.00  \\
   &   &     &  8.00  & -10.84   & -10.84 & 0.00   \\
   &   &     &  12.00  &  -0.945  & -0.946 & 0.11 \\
   &   &     &  14.00  &  -0.349  & -0.349 & 0.00  \\
\end{tabular}
\end{table}

The interaction potential $V$ is written using a spherical harmonics expansion 
as\cite{Wormer:84}:

\begin{equation}
V(R,\theta_a,\theta_b,\phi) \, = \, (4\pi)^{3/2} \, \sum_{l_a,l_b,l} \, f^{l_a l_b l}(R) 
\, A_{l_a l_b l}(\theta_a,\theta_b,\phi),
\label{expansion}
\end{equation}

\noindent
with

\begin{eqnarray}
A_{l_a l_b l} (\theta_a,\theta_b,\phi) \,&  = & \, \left( \frac{2 l + 1}
{4 \pi}\right)^{1/2} \, \sum_{m} \, 
\left( \begin{array}{rrr} l_{a} & l_{b} & l \\ m & -m & 0 \end{array} \right) \nonumber \\ 
 & & Y_{l_a,m}(\theta_a,0) \, Y_{l_b,-m}(\theta_b,\phi),
\label{sphericalhar}
\end{eqnarray}

\noindent
where $Y_{l_a,m}$ and $Y_{l_b,-m}$ are the normalized spherical harmonics that are coupled
with a 3$j$ symbol, and $m$ runs from $-\min(l_a,l_b)$ to
$\min(l_a,l_b)$. The radial coefficients $f^{l_a l_b l}(R)$ are computed 
by integrating $V$ over the angular variables:

\begin{eqnarray}
f^{l_a l_b l}(R) &  = & \pi^{1/2} \int_{-1}^1 d(\cos\theta_a) \int_{-1}^1 d(\cos\theta_b) \int_0^{2\pi} d\phi \nonumber \\
& & V(R,\theta_a,\theta_b,\phi) A^{*}_{l_a,l_b,l} (\theta_a,\theta_b,\phi).
\label{radcoeff}
\end{eqnarray}

\noindent
Integrals were carried out for each intermolecular distance $R$ by means of
Gauss-Legendre (for $cos\theta_{a,b}$) and Gauss-Chebyshev (for $\phi$) 
quadratures\cite{abramovitz}.

 Due to the permutation-inversion symmetries of the complex, $l_a$, $l_b$,
  and $l$ integers must be even. In contrast with dimers of identical
 diatoms, however, $f^{l_a l_b l}$ and $f^{l_b l_a l}$ coefficients 
 differ. From the set of quadrature points used in the present work, a series of 85 different
 radial coefficients arise, consisting in all the combinations $(l_a l_b l)$ where
 $l_a, l_b$= 0, ..., 8  and  $\left| l_a-l_b \right|$ $\leq l \leq$
 $l_a$+$l_b$. We have selected a subset of 39 radial coefficients for which
 the interaction potential is sufficiently well represented. It was built by
 discarding, for $l_a > 4$ or $l_b > 4$, all coefficients with $l \neq l_a+l_b$, 
 except for $(l_a l_b l)$= (2 6 4),  (2 6 6), (6 2 4) and (6 2 6). Root
 mean square (rms) relative and absolute errors of this expansion with respect
 to the {\em ab initio} energies (for all distances and quadrature points) are
 about 2 \% and 0.55 meV, respectively.

A further indication of the quality of the spherical harmonics expansion is 
given in Table \ref{table2}, where a comparison between {\em ab initio}
interaction energies and those corresponding to the expansion is given 
for some selected geometries. This is an independent check of the
accuracy of the expansion, since, except for the {\sf X} geometry, 
none of these geometries correspond to the Gaussian quadrature 
points used to build the PES and then, they were computed independently. 
It can be seen that the comparison is satisfactory. The largest discrepancies
are found at short intermolecular distances ($R  \leq  5.5$ bohr) where
the interaction becomes more anisotropic.

\subsection{Long range behavior and matching procedure}

\begin{figure}[t]
\includegraphics[width=9cm,angle=0.]{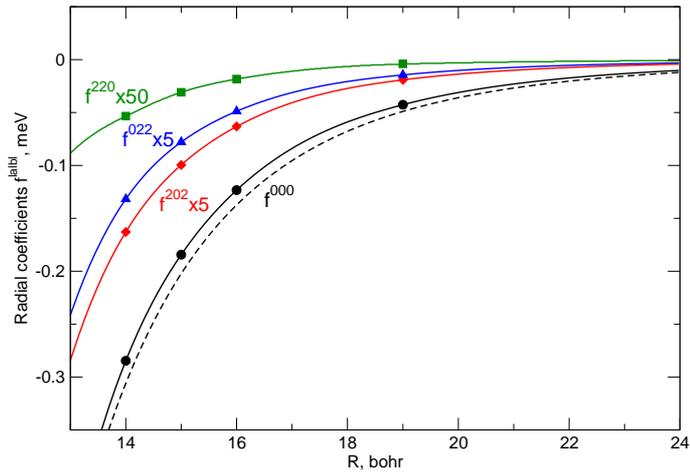}
\vspace{1.cm}
\caption[]{Behavior of the $f^{000}$, $f^{202}$, $f^{022}$ and $f^{220}$ radial
coefficients at large intermolecular distances. Points at 14, 15 and 16 bohr
are those resulting from the {\em ab initio} calculations, whereas Eq. (\ref{lr})
is used for $R$= 19 bohr. For $R<$ 19 bohr the curves are the result of cubic spline
interpolation, while for larger $R$'s they are given by the analytical long
range  behavior of Eq. \ref{lr}. The dashed line corresponds to the $f^{000}$
term of the experimentally derived PES of Ref.\cite{Aquilanti:03}.}
\label{fig2}
\end{figure}

In this paragraph we present the asymptotic extension of the PES by using
analytical expressions for the long range behavior. Following the 
Rayleigh-Schr\"odinger perturbation
theory of intermolecular forces \cite{Cwiok:92} in the multipole 
approximation \cite{Avoird:80,Heijmen:96}, 
the radial coefficients $f^{l_a l_b l}(R)$ of 
Eq. \ref{expansion} can be written by a sum of electrostatic ($el$),
dispersion ($d$) and induction ($i$) contributions: 

\begin{equation}
f^{l_a l_b l}(R) = f^{l_a l_b l}_{el}(R) +  f^{l_a l_b l}_{d}(R) +  
f^{l_a l_b l}_{i}(R),
\label{lr}
\end{equation}

\noindent
where 

\begin{equation}
f^{l_a l_b l}_{el}(R) = \delta_{l_a+l_b,l} \left[
\frac{(2l_a+2l_b)!}{(2l_a+1)!(2l_b+1)!} \right]^{1/2}
\frac{Q_0^{l_a}Q_0^{l_b}}{R^{l_a+l_b+1}},
\label{lrel}
\end{equation}

\noindent
and

\begin{equation}
f^{l_a l_b l}_{(d,i)}(R) = -\frac{1}{[(2l_a+1)(2l_b+1)(2l+1)]^{1/2}} 
\sum_n \frac{C_{n,(d,i)}^{l_a l_b l}}{R^{n}}.
\label{lrd}
\end{equation}

Coefficients involved in Eqs. \ref{lrel} and \ref{lrd} have recently been
obtained from high level {\em ab initio} calculations\cite{max-lr:10}. 
Permanent electric multipole moments $Q_0^{l}$ ($l$=2,4,6,8) were taken
from Tables III and IV of Ref.\cite{max-lr:10}, and dispersion and induction 
coefficients $C_{n,(d,i)}^{l_a l_b l}$, (with $n$=6,8 and $n$=8, respectively, 
and $(l_a l_b l)$ up to (2 4 6) ), from Table VII of the same work.

The expression of Eq. \ref{lr} was used for distances $R \ge 19$
bohr. For shorter distances, $f^{(l_a l_b  l)}(R)$ were calculated from cubic
spline interpolation of the values $f_i^{(l_a l_b l)}$ obtained from the
SAPT(DFT) calculations at the set of distances
$R_i, i=1, ..., N$. An additional point $R_{N+1}$=19 bohr was
included in that grid with a value given by Eq. \ref{lr}. 
The analytical derivative of $f^{l_a l_b l}(R)$ at $R_{N+1}$
was used as a boundary condition for solving the interpolation equations, so
that a smooth behavior of the radial term around $R_{N+1}$ is achieved. 
The procedure worked well, providing a good indication of the
consistency between the SAPT(DFT) calculations and the computed electric
properties of the molecular fragments.  As an example, we show 
in Fig.\ref{fig2} the behavior of the first four coefficients around the
matching region. In there the long range behavior of 
$f^{000}$ from the Perugia-PES\cite{Aquilanti:03}
is also reported.

\section{Results and discussion}
\label{results}

\subsection{The potential energy surface}

\begin{figure}[t]
\includegraphics[width=8cm,angle=0.]{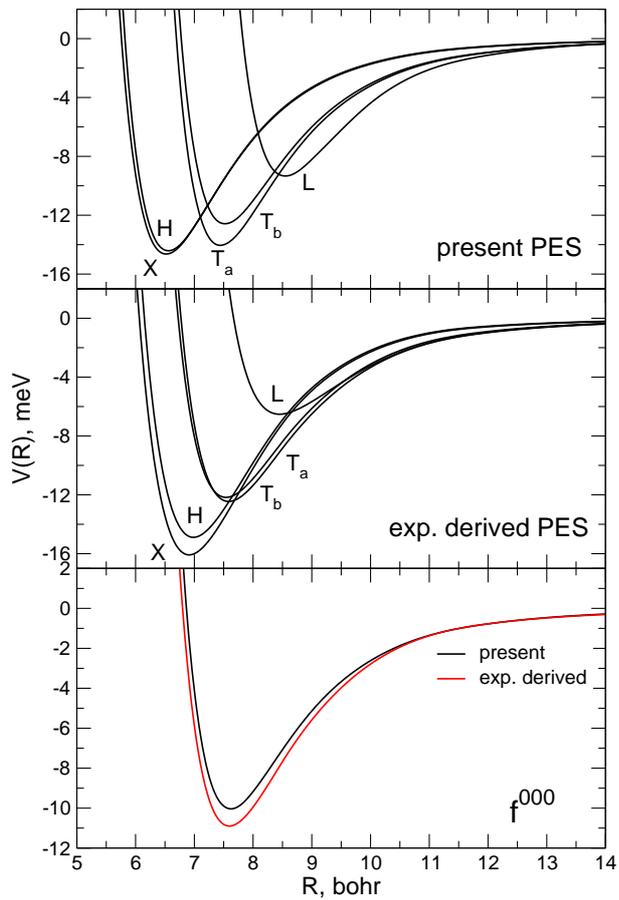}
\caption[]{Comparison between present {\em ab initio} PES and the
  experimentally derived one from Ref. \cite{Aquilanti:03}. Upper and
  intermediate panels: potential energy (in meV) profiles as functions of the
  intermolecular distance $R$ (in bohr) for selected geometries of the
  dimer. Lower panel: spherical average $f^{000}$ as a function of $R$.} 
 \label{fig3}
\end{figure}

\noindent
\begin{table}
\caption{Equilibrium distances $R_m$ (in bohr) and well depths $D_m$ (in meV)
  of the O$_2$-N$_2$ PES at some selected orientations as well as the spherical
  average. Present results are compared with those of 
Perugia-PES\cite{Aquilanti:03}.} 
\vspace{0.3cm}
\vspace{0.1cm}
\tabcolsep0.5cm
\begin{tabular}{lrrrr}
\hline
  & \multicolumn{2}{c}{Ref.\cite{Aquilanti:03}} &  \multicolumn{2}{c}{This work}\\
\hline 
 & $D_m$  & $R_m$  & $D_m$  & $R_m$ \\
\hline
{\sf H} & 14.89 & 6.99 &   14.41        & 6.56 \\
{\sf X} & 16.08 & 6.92 &   14.63        & 6.52 \\
{\sf T$_a$} & 12.17 & 7.54 & 14.05      & 7.44 \\
{\sf T$_b$} & 12.45 & 7.58 & 12.58      & 7.52 \\
{\sf L} & 6.52 & 8.47      &  9.33     &  8.54\\
{\sf S$_{45}$} & 8.76 & 8.01 &  11.68      & 7.66 \\
f$^{000}$ & 10.90 & 7.60     &   10.04    & 7.62  \\
\hline
\end{tabular}
\label{table3}
\end{table}

\begin{figure}[t]
\includegraphics[width=8cm,angle=0.]{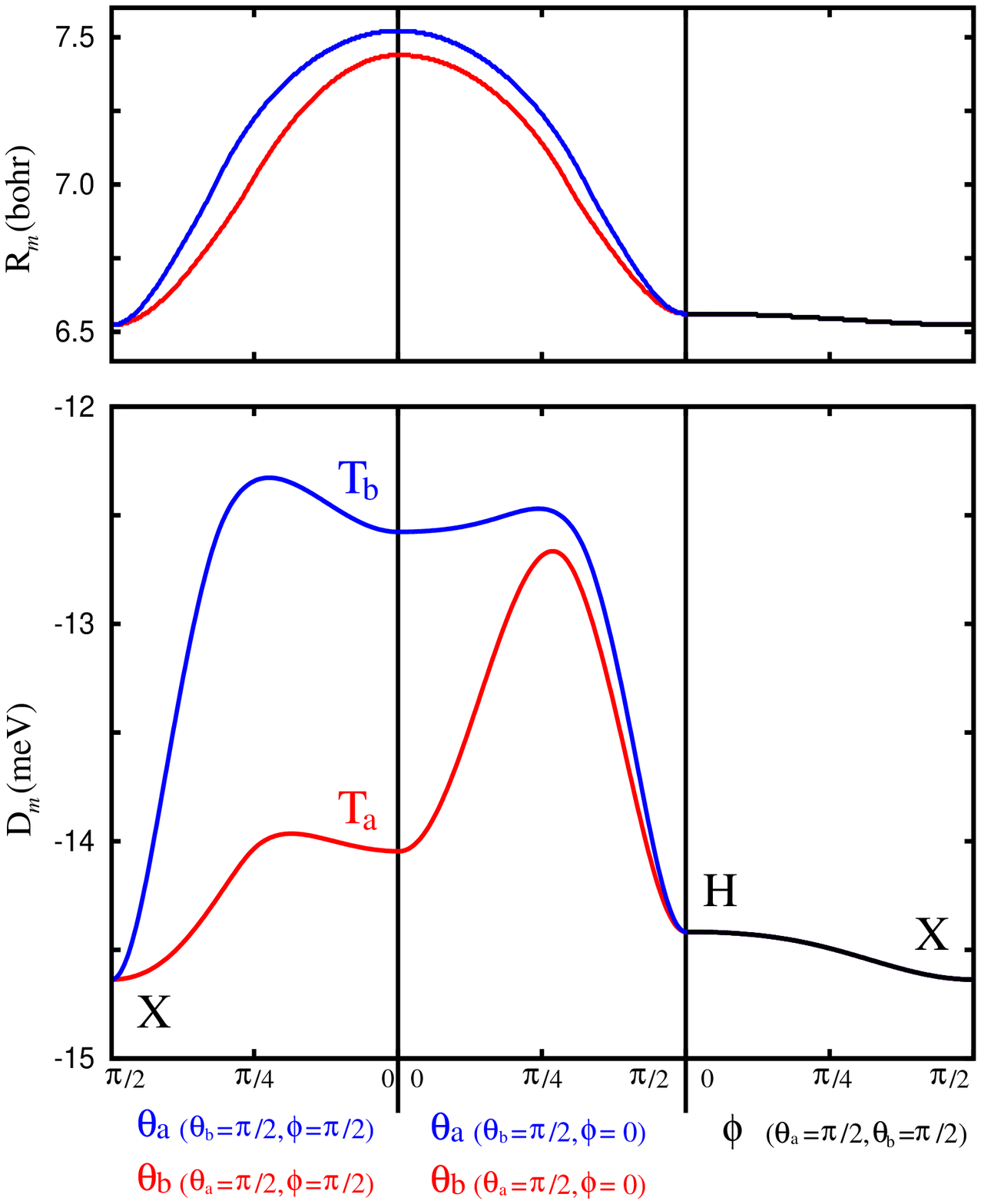}
\caption[]{Low energy pathways of the present O$_2$-N$_2$ PES. As the relative
  orientation of the diatoms is varied as indicated in the abscissae labels, the
  associated equilibrium distances $R_m$ (upper panel) and well
depths $D_m$ (lower panel) are displayed. Left panels: 
  from the crossed {\sf X} to {\sf T$_a$} and {\sf T$_b$} geometries; center
  panels: from the latter to the {\sf H} configuration;  right panels: from
  {\sf H} to the {\sf X} orientation.}
 \label{fig4}
\end{figure}

In Fig. \ref{fig3} and Table \ref{table3}, we report the most prominent
features of the present PES in comparison with those of the 
Perugia-PES\cite{Aquilanti:03}. Both intermolecular potentials are in a good
qualitative agreement. For example, the crossed configuration ({\sf X}) is the
most stable geometry for both PESs, while a good match in the  equilibrium
distance and well depth of the spherically averaged interaction is observed. 
Nevertheless, some important differences can be found when going into details.
In fact, present calculations give shorter equilibrium distances for
the parallel ({\sf H}) and the crossed ({\sf X}) configurations, the curves
being almost identical for these two orientations. Moreover, the {\em ab
  initio} well depth of the {\sf T$_a$} geometry is larger than the {\sf
  T$_b$} one, and it is very close to those of the most stable {\sf X} and {\sf
  H} arrangements. This is in contrast with the features of the T-shaped
geometries of the Perugia-PES\cite{Aquilanti:03} (see the intermediate panel of
Fig. \ref{fig3}). In general, the present PES
provides deeper potentials for the more repulsive geometries (T-shaped and
collinear {\sf L}), while for the most attractive ones ({\sf X} and {\sf H}), 
slightly shallower wells. In spite of this, we believe that the {\em ab
initio} PES is more anisotropic since, at a given $R$, the
interaction energy exhibits a stronger dependence with the angles than in the
Perugia-PES case.

The {\em ab initio} spherically averaged radial term ($f^{000}$) is compared
with that of Ref.\cite{Aquilanti:03} in the lower panel of Fig. \ref{fig3}.
This test is meaningful because the latter was obtained from analyses of
accurate integral cross section data\cite{Brunetti:81}. As can be seen,
although present calculations slightly underestimate the experimentally derived
interaction energy in the well and repulsive regions, the matching for the
equilibrium distance, $R_m$, is very good (see Table \ref{table3}). Differences
in the interaction energy are less than 1 meV near the equilibrium distance
and become almost negligible in the long range tail (see also
Fig. \ref{fig2}).

More insight into the features of the O$_2$-N$_2$ interaction is gained from
Fig. \ref{fig4}, where we show a few representative pathways joining the most
stable configurations {\sf X, T$_a$, T$_a$}, and {\sf H}. First, it is
worthwhile noticing the  almost flat path between the {\sf H} and {\sf X}
geometries, suggesting a free internal (torsional) rotation in the dimer. In
addition, it can be seen that the paths connecting the T-shaped structures
with the most stable {\sf H} or {\sf X} geometries involve very small barriers 
(except for the {\sf T$_a$}-{\sf H} path). This finding 
brings a picture of the O$_2$-N$_2$ dimer as a rather floppy cluster.

\subsection{Integral cross sections}

\begin{figure}[t]
\includegraphics[width=8.cm,angle=0.]{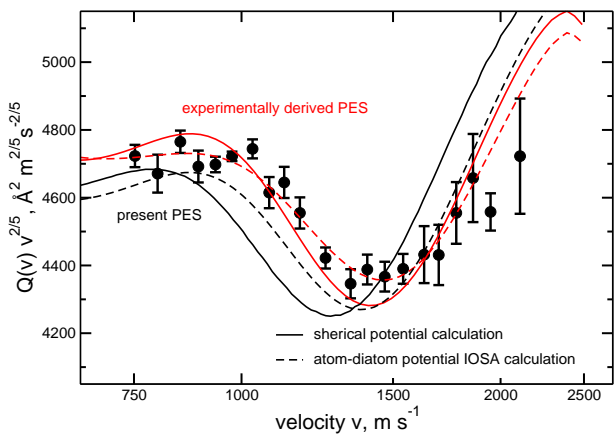}
\caption[]{Total integral cross sections $Q(v)$ times $v^{2/5}$, for
  scattering of a rotationally hot O$_2$ effusive beam by  N$_2$ target
  molecules as functions of the beam velocity, $v$. Present calculations are
  in solid and dashed black lines as obtained with a spherical average and an
  atom-diatom dynamical models, respectively (see text). 
Experimental data (filled circles with error bars) and related best fit
calculations (in red) are from Refs. \cite{Brunetti:81,Aquilanti:03}.
}
\label{fig5}
\end{figure}

In order to assess the reliability of the present PES, 
we have tested it against available experimental data such as the total
integral cross sections reported in Refs. \cite{Brunetti:81,Aquilanti:03} for
the scattering of rotationally hot and near effusive O$_2$ beams. In these
experiments, oxygen projectiles are at a high rotational temperature (about
10$^3$ K) and collide with target N$_2$ molecules which are in the scattering
chamber at low translational temperature.  Under these conditions Brunetti
et al. \cite{Brunetti:81} were able to resolve the glory structure which
in turn can give valuable
information\cite{cappelletti:08,pccpnon2:08,jpcac2h2:09} on the  
intermolecular interaction involved. As explained before (see
Ref. \cite{Aquilanti:03} and references therein), at low collision velocities, the
colliders mainly probe the isotropic component of the interaction,  while the
anisotropy plays an increasing role at intermediate and high  velocities.

Accordingly, we have computed total cross sections $Q(v)$ using two different
schemes. In one case, we have used just the first term of the expansion of
Eq. \ref{expansion}, $f^{000}$, which represents the spherical average of the
PES. In a second approach, we have used the potential energy
surface in an atom-diatom limit, corresponding to the interaction averaged over
the orientation of the O$_2$ projectile\cite{Aquilanti:03}. In the latter case,
the surface reduces to the sum of just five terms arising in
Eq. \ref{expansion} when the $(l_a,l_b,l)$= (0 $l_b$ $l_b$) constraint is
imposed  ($l_b$=0,2,4,6,8) and the calculations where performed using
the infinite-order-sudden (IOS) approximation.  In both approaches the phase
shifts have been evaluated within the JWKB approximation.
The center-of-mass cross sections are then convoluted over
the relative velocity distribution and transformed to the laboratory reference
frame.  More technical details can be found in Ref. \cite{jpc-benzenehot:02} 
and references therein.

In Fig. \ref{fig5} we report the cross sections obtained within the
two proposed schemes and using the {\em ab initio} PESs, in comparison with
those obtained using the experimentally derived PES as well as the
experimental results\cite{Aquilanti:03}. An overall good agreement is
noticed. 
The average value of the {\em ab initio} cross sections is just a few percents
lower ($<$ 2 \%) than the best experimental fit, with the glory pattern
slightly shifted towards lower velocities. The IOS results with the 
effective atom-diatom potential
better describe the observed glory oscillations than the isotropic potential. 
It is generally admitted that
the average values of the cross sections contain information about the
long-range attraction\cite{schiff56,Pirani:82}.  In this context and for the
velocities probed experimentally, the ``long range region'' corresponds to
intermolecular distances of about 11-14 bohr\cite{o2kr,Jesus:09} where a good
matching between the present and the experimentally derived PESs was already
noted (see Figs. \ref{fig2} and \ref{fig3}). On the other hand, it is also well known
that the glory structures of the cross sections provide information on the
potential well area\cite{Bernstein:65,Bernstein:67}. In addition to this, it
has been pointed out recently\cite{jpr-o2o2-cpl-12} that anisotropy can also
influence the velocity positions of the
glory extrema in molecule-molecule scattering. In fact, it can be seen in
Fig.\ref{fig5} that the glory extrema shift to higher velocities when one
moves from the atom-atom to the atom-diatom approximation. Thus, we
expect that the full molecule-molecule scattering calculation would shift the
glory pattern towards even higher velocities (as found for
O$_2$-O$_2$\cite{jpr-o2o2-cpl-12}), eventually bringing theoretical results to a 
closer agreement with the measurements. 

\subsection{Second virial coefficients}

\begin{figure}[t]
\includegraphics[width=8.cm,angle=0.]{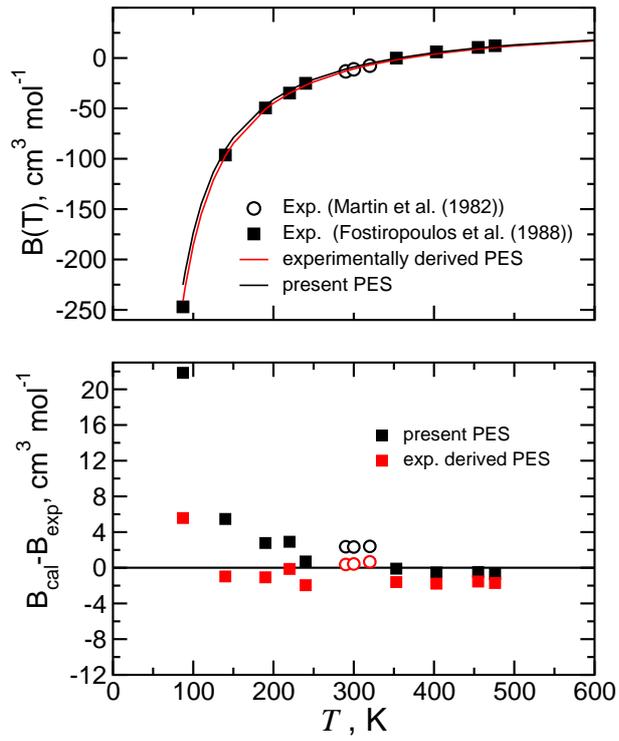}
\vspace*{.2cm}
\caption[]{Upper panel: Second virial coefficients $B(T)$ as a function of
  temperature. Present calculations are compared with those obtained with the
  Perugia-PES\cite{Aquilanti:03} as well as with the
  experimental data of Martin {\em et al }\cite{Dunlop:82} and Fostiropoulos
  {\em et al} \cite{n2o2vir}. Lower panel: Deviations
  between experimental and calculated $B(T)$'s. 
}
\label{fig6}
\end{figure}

\begin{table}
\vspace*{-.5cm}
\footnotesize
\renewcommand{\baselinestretch}{0.7}
\caption{Calculated and measured values of the second virial coefficient $B(T)$
(cm$^{3}$ mol$^{-1}$)  as a function of temperature (in K). Values
corresponding to the contribution given by the spherical component of the
interaction are shown in parentheses.}
\label{tablevir}
\renewcommand{\arraystretch}{0.8}
\tabcolsep0.15cm
\begin{center}
\begin{tabular}{rrrrrr}
\hline
$T$ (K) & \multicolumn{2}{c}{PES Ref.\cite{Aquilanti:03}} &
\multicolumn{2}{c}{This work} & Experimental  \\
\hline
87.2   & -241.4 & (-206.1) & -225.2 & (-172.1)  & -247.0$\pm7$$^a$   \\
90.0   & -227.1 & (-194.2) & -212.0 & (-162.1)  &          \\
100.0   & -185.7 & (-159.3) & -173.5 & (-132.5)  &         \\
110.0   & -155.0 & (-133.0 ) & -144.9 & (-110.0)  &       \\
125.0   & -121.4 & (-103.8)  & -113.5 & (-84.9)  &          \\
140.0  & -97.3  & (-82.7)   & -90.8 &  (-66.6) & -96.3$\pm5$$^a$    \\
190.0  & -50.7  & (-41.2) &  -46.8  & (-30.2) & -49.6$\pm4$$^a$    \\
220.0  &  -34.9 & (-27.0) &   -31.9 & (-17.6) &  -34.8$\pm3$$^a$     \\
240.0  &  -27.1 & (-19.8) &   -24.4 & (-11.3) & -25.1$\pm2$$^a$     \\
290.0  &  -12.9 & (-6.9) &  -10.9 & (0.1) &        -13.3$^b$  \\
300.0  &  -10.9 & (-5.0) & -8.9 & (1.9) &        -11.2$^b$  \\
320.0  &  -6.9 & (-1.5) & -5.2 & (5.0) &    -7.6$^b$  \\
353.0  & -1.7 & (3.3) & -0.2 & (9.3)  &  -0.1$\pm2$$^a$       \\
403.0  &  4.3 & (8.9) & 5.6  & (14.2)  &   6.1$\pm3$$^a$      \\
455.0  &  8.9 & (13.1) & 10.0 & (18.0)  &   10.5$\pm3$$^a$       \\
476.0  &   10.5 & (14.5) & 11.5 & (19.2) &   12.2$\pm3$$^a$      \\
500.0  &   12.1 & (16.0) & 13.0 & (20.5) &         \\
600.0  &   17.0 & (20.5) & 17.8 & (24.5) &         \\
700.0  &   20.3 & (23.5) &  21.0 & (27.2) &         \\
800.0 &   22.6 & (25.6) & 23.2 & (29.0) &           \\
900.0  &  24.2  & (27.0) & 24.8 & (30.2) &         \\
1000.0 &   25.4 & (28.1) & 26.0 & (31.2) &         \\
\hline
\end{tabular}
\end{center}
\begin{flushleft}
\begin{footnotesize}{
$^a$ Fostiropoulos {\em et al}\cite{n2o2vir}\\
$^b$ Martin {\em et al }\cite{Dunlop:82}\\}
\end{footnotesize}
\end{flushleft}
\end{table}

A further check of the quality of the present PES is carried out through
the computation of the second virial coefficient $B(T)$ as a function of the
temperature $T$. To this end we apply the semiclassical theory based on the
expressions for two linear molecules presented by Pack\cite{Pack:83} which
include the first quantum correction due to the relative translational and 
rotational motions, including Coriolis coupling.  

In Table \ref{tablevir} and in Fig. \ref{fig6} (upper panel), the calculated
values of $B(T)$ are compared with the experimental data of
Refs.\cite{Dunlop:82,n2o2vir}, together with those corresponding to the
Perugia-PES\cite{Aquilanti:03}. A more detailed analysis of the
deviations of the calculations with respect to the measurements is given in the
lower panel of Fig. \ref{fig6}. It can be seen that, except for the results at
$T$= 87.2 K, the comparison with the experimental data of Ref.\cite{n2o2vir}
is very good and almost perfect around and above the Boyle temperature
($T\sim$ 353 K). In fact, the {\em ab initio} PES works better in the high
temperature region than the experimentally derived PES. These results indicate
that the present PES is well characterized, especially in the repulsive
region.  

The role of the anisotropy can be inferred from the comparison of $B(T)$ 
obtained using the full PES with a calculation where the
anisotropy is neglected just by retaining the isotropic component of the
interaction. The results of the latter calculations are shown in parentheses in
Table \ref{tablevir}. It can be noticed that there are significant differences
between the two approaches and that this discrepancy increases as temperature
decreases. Interestingly, the isotropic approximation as obtained using the
{\em ab initio} PES leads to less negative and more
positive $B(T)$ at low and high temperatures, respectively, if compared with
the results from the experimentally derived PES. This is due to the less attractive
 character of the isotropic component for the {\em ab initio} PES as already 
 discussed above (see also Fig.\ref{fig3}). However, 
 the contribution of the anisotropy is more important in
 the {\em ab initio} than in the experimentally derived PES, which in turn
 compensates the isotropic contribution and  globally leads to  a  good
 agreement with the experimental $B(T)$. 
 
Since second virial coefficient measurements for gas mixtures are in general
less accurate than those for pure gases \cite{virmix:90} we find useful to
provide data for temperatures higher than those considered in the experiments,
as shown in Table \ref{tablevir}. Based on the previous analysis, we believe
that present results for the second virial coefficients $B(T)$ are the best 
theoretical estimates to date for
temperatures higher than 353 K and, for $T>$  500 K, they should be taken as
the reference data.

\section{ Bound state calculations }
\label{bound}

\begin{table}
\caption{Energies (in meV and cm$^{-1}$) of the lowest rotationless
  ($J$=0) states of $^{16}$O$_2$-$^{14}$N$_2$. 
The parity of $j_a$ (O$_2$), $j_b$ (N$_2$) and $l$ in the wave function
expansion is indicated, together with the parity of the total  
eigenfunctions (including the electronic part) with respect to the operations
 of the $G_{8}$ group, and their symmetry (as in Table A-22, page 393 of
 Ref.\cite{Bunker-book}). Allowed states are given in boldface.}
\label{table5}
\begin{tabular}{c r r c c c c c c}
\hline
Level  & $E$(meV) & $E$(cm$^{-1}$) & & $j_a \; j_b \; l$ &  
                    & $P_{12} \; P_{34} \; E^*$ & & Sym \\
\hline
 0 & -10.32 & -83.28 & & $e \; e \; e$ & & $- \; + \; +$ &  & B$_2''$ \\
{\bf 1} & {\bf -9.81} & {\bf -79.11} & & {\bf $o \; o \; o$} & & {\bf $+ \; -
  \; -$} & & {\bf B$_1'$} \\ 
{\bf 2} & {\bf -9.72} & {\bf -78.37} & & {\bf $o \; o \; e$} & & {\bf $+ \; -
  \; +$} & & {\bf B$_2'$} \\ 
 3 & -9.39 & -75.78 & & $e \; o \; o$ & & $- \; - \; +$ & & A$_1''$ \\
{\bf 4} & {\bf -9.16} & {\bf -73.89} & & {\bf $o \; e \; o$} & & {\bf $+ \; +
  \; +$} &  & {\bf A$_1'$} \\
 5 & -8.75 & -70.59 & & $e \; e \; e$ & & $- \; + \; +$&  & B$_2''$ \\
 6 & -8.65 & -69.79 & & $e \; e \; e$ & & $- \; + \; +$&  & B$_2''$ \\
{\bf 7} & {\bf -8.33} & {\bf -67.23} & & {\bf $o \; e \; e$} & & {\bf $+ \; +
  \; -$} &  & {\bf A$_2'$} \\
{\bf 8} & {\bf -8.32} & {\bf -67.09} & & {\bf $o \; e \; o$} & & {\bf $+ \; +
  \; +$} &  & {\bf A$_1'$} \\
 9 & -8.25 & -66.53 & & $e \; e \; o$ & & $- \; + \; -$&  & B$_1''$ \\
 10 &  -8.24 & -66.46 & &  $e \; e \; e$ & & $- \; + \; +$ & &  B$_2''$ \\ 
 11 &  -8.19 & -66.08 & &  $e \; o \; o$ & & $- \; - \; +$ & &  A$_1''$ \\ 
 12 &  -8.14 & -65.68 & &  $e \; o \; e$ & & $- \; - \; -$ & &  A$_2''$ \\ 
\hline
\end{tabular}
\end{table}

In this section, we report energies and geometries of the lowest bound 
states of O$_2$-N$_2$. Since the intramolecular distances have been fixed in
the {\em ab initio} PES, the diatoms O$_2$ and N$_2$ are treated as
rigid rotors wihin the present approach. The Hamiltonian describing the
nuclear motion writes -using the previously defined diatom-diatom
Jacobi coordinates in a space-fixed (SF) frame as \cite{Green75,Avoird:94} 

\vspace{-.2cm}

\BEA
 H& =& -\frac{1}{2 \mu R} \frac{\partial^2}{\partial R^2} R +
\frac{{\bf \hat {\ell}} ^2}{2 \mu R^2} + B_a \, {\bf{\hat{j}_a}}^2 + 
B_b \, {\bf{\hat{j}_b}}^2 + V,
\label{Hamil}
\EEA

\noindent
where atomic units are used ($\hbar=1$), $\mu$ is the reduced mass of the
complex, $V$ is the PES, $B_{a,b}$ are the rotational
constants of O$_2$ and N$_2$, respectively, and ${\bf \hat{\ell}}$,
${\bf{\hat{j}_a}}$ and ${\bf {\hat{j}_b}}$ are the angular momentum operators
associated with the rotations of the  ${\bf R}$, ${\bf r}_a$, and  ${\bf r}_b$ vectors,
respectively. In this way, the total angular momentum is 
${\bf J}= {\bf \hat{\ell}} + {\bf{\hat{j}_a}} + {\bf {\hat{j}_b}}$. 

Bound states of the complex were obtained by solving the time-independent
Schr{\"o}dinger equation, as described in Ref.\cite{Hutson94}. 
Eigenfunctions with well defined values of $J$ and $M$ (projection onto the SF
$z$ axis) are sought by using a convenient basis set expansion 

\vspace{-.2cm}

\BEA
\nonumber
\Psi^{JM} ({\bf{R},\bf{\hat r_a},\bf{\hat r_b}})\hspace{-.1cm} & = & 
\hspace{-.3cm}
\sum_{j_a j_b j_{ab} \ell} 
\hspace{-.2cm} { g^{J M}_{j_a j_b j_{ab} \ell} (R) \over R}
I^{J M}_{j_a j_b j_{ab} \ell}(\bf{\hat R},\bf{\hat r_a},\bf{\hat r_b}),\\
 & & 
\label{expfi}
\EEA

\vspace{-.2cm}

\noindent
where ${\bf {\hat R}, \bf{\hat r_a}, \bf{\hat r_b}}$ are unit vectors (i.e.,
${\bf \hat {R}} = {\bf R}/R$, etc.) and the angular wavefunctions
are\cite{Green75,Alexander,Avoird:94} 

\begin{equation}
I^{J M}_{j_a j_b j_{ab} \ell}({\bf{\hat R}},{\bf{\hat r_a}},{\bf{\hat r_b}}) =
\left[ [Y_{j_a}({\bf{\hat r_a}})\otimes Y_{j_b}({\bf{\hat r_b}})]^{j_{ab}}
\otimes Y_l({\bf{\hat R}})\right]^J_M
\label{funang}
\end{equation}

\vspace{-.2cm}

\noindent
where 
the irreducible tensor product of two
spherical tensors is given by
\begin{equation}
\left[T_l\otimes T_{l'}\right]_L^K =
\sum_{k=-l}^l\sum_{k'=-l'}^{l'}\langle l,k;l',k'|L,K\rangle T_l^k T_{l'}^{k'},
\label{tp}
\end{equation}
and $\langle l,k;l',k'|L,K\rangle$ is the Clebsch-Gordan coefficient.

The molecular symmetry group appropriate for describing the 
 O$_2$-N$_2$ dimer is $G_{8}$, a permutation-inversion group generated by the
 spatial inversion of the nuclear coordinates, $E^*$,  and the
permutations of identical nuclei within O$_2$ and N$_2$, $P_{12}$ and
$P_{34}$, respectively. The basis functions of Eq.\ref{funang} are well
adapted to these operations, with parities given by  $(-1)^{j_a+j_b+\ell
}$\cite{Alexander},  $(-1)^{j_a}$ and $(-1)^{j_b}$ for $E^*$,
$P_{12}$ and $P_{34}$, respectively. To complete the analysis, we have to
consider the effect of 
these operations on the electronic wave function since it depends
parametrically on the coordinates of the nuclei. This function,
correlating with the $^3\Sigma^-_g + ^1\Sigma^+_g$ limit, is symmetric under all the
group operations except for the exchange of the oxygen nuclei, $P_{12}$, for which
it is antisymmetric \cite{Carrington-book}. Thus, the parity of the {\em total}
wavefunction under $P_{12}$ is  $(-1)^{j_a+1}$. In addition, calculations
have been carried out for the $^{16}O$ 
nuclei, which have zero spin. Therefore, the total wave function must be
symmetric under $P_{12}$ and states with even $j_a$ are forbidden. 

Calculations were performed with the {\sf BOUND} package\cite{bound}.
The coupled differential equations resulting from substitution of
Eq. \ref{expfi} into the Schr{\"o}dinger equation were propagated  
from 5.3 to 12.9 bohr, with a step size of 0.015 bohr using the Manolopoulos
modified log-derivative algorithm \cite{Mano}, and eigenvalues were located
iteratively by using the Johnsons's method\cite {Johnson}. The most abundant $^{16}$O
and $^{14}$N isotopes are considered, hence the reduced mass is 14.936928
amu. The rotational constants  
$B_a$ and $B_b$ are 1.438 and 2.013 cm$^{-1}$, respectively. An angular basis
setwith $j_a$,  $j_b$ $\leq$ 14 was used. Note that we
have computed bound states for {\em all} the symmetries including forbidden
states with even $j_a$, so that a more detailed understanding of the
rovibrational structure of the complex can be achieved. The resulting energy
levels are converged to within 10$^{-4}$ cm$^{-1}$ at worst.

\begin{figure}[t]
\includegraphics[width=7.5cm,angle=0.]{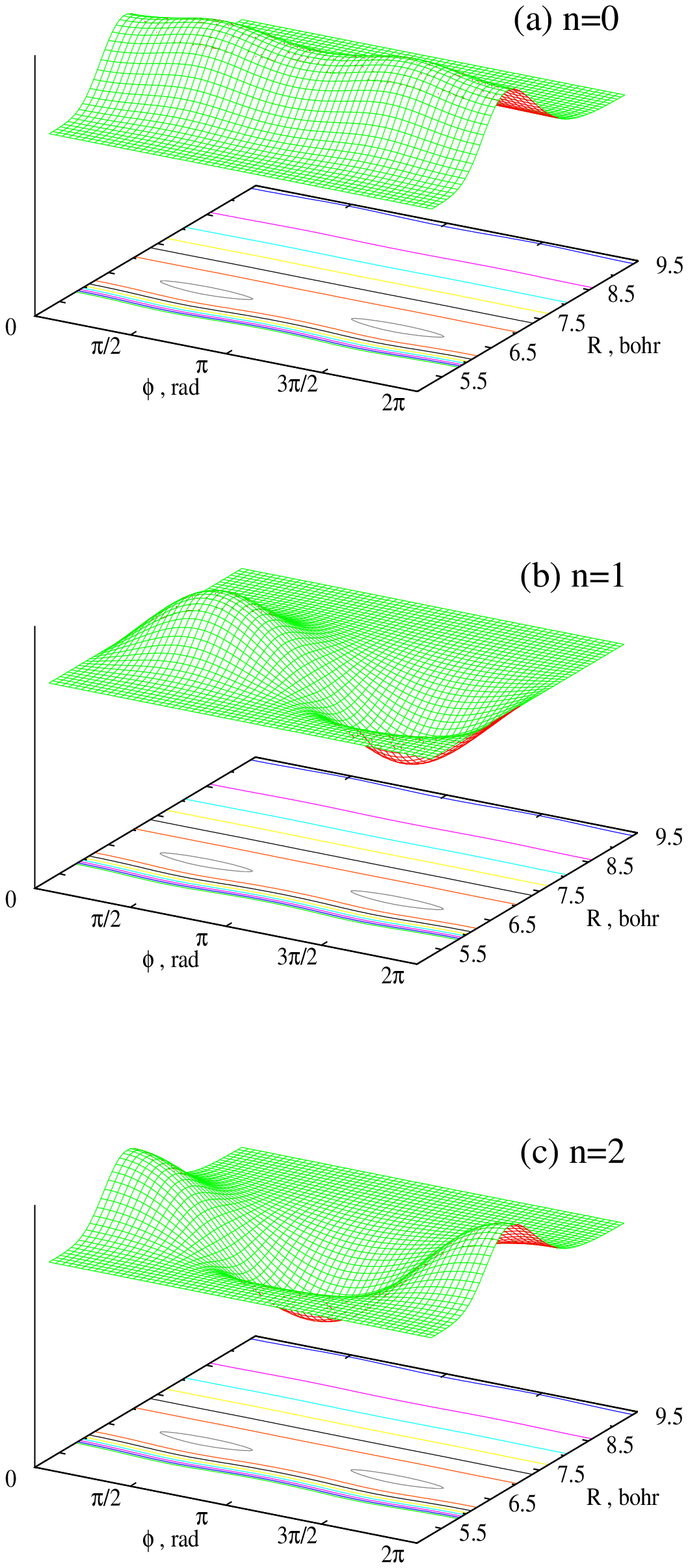}
\caption[]{Wave functions multiplied by $R$ (in arbitrary units)
  for the lowest  vibrational
  states of O$_2$-N$_2$, as functions of the intermolecular distance $R$
  (bohr) and the torsional angle (radian). Contour plots of
  the PES are also shown in the bottom planes of the plots, with contour lines
  ranging from -14.5 to -0.5 meV, in steps of 2 meV. In all cases, 
the  bending angles $\theta_a$ and  $\theta_b$  have both been fixed at $\pi/2$. 
 (a) $n=0$, real part, (b) $n=1$, imaginary part (c) $n=2$, real part; (the
  associated imaginary/real parts are zero).  
}
\label{fig7}
\end{figure}

\begin{figure}[t]
\includegraphics[width=7.5cm,angle=0.]{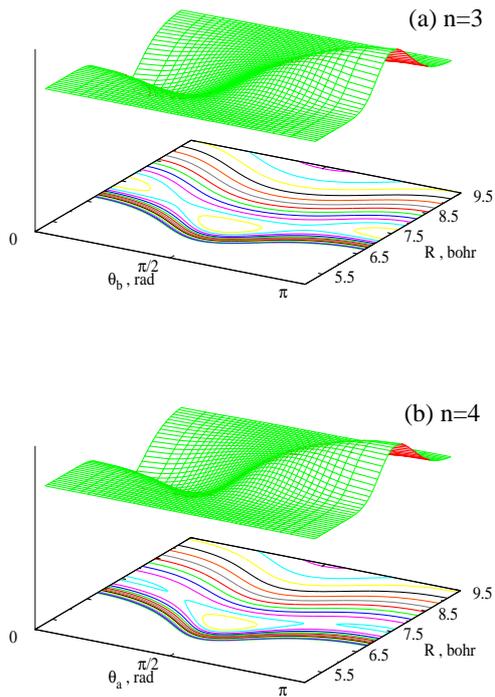}
\caption[]{(a) Real part of the wave function multiplied by $R$ 
(in arbitrary  units) of
  the third excited rotationless 
  state ($n=$3) of O$_2$-N$_2$, as a function of $R$ (bohr) and $\theta_b$
  (radian), with the bending angle $\theta_a$ fixed at $\pi/2$ and $\phi=$
  0. Imaginary part is zero. (b) As in (a), for $n=$4. 
  In this case the wave function is shown versus $R$ and $\theta_a$, 
  with  $\theta_b = \pi/2$ and $\phi=$ 0. In this way, these states
  correspond to $T_a$ and $T_b$ configurations, respectively.  Contour plots of
  the PES are also shown in the bottom planes of the plots, with contour lines
  ranging from -13.5 to -0.5 meV, in steps of 1 meV. 
}
\label{fig8}
\end{figure}

\begin{table}
\caption{Energies (in cm$^{-1}$) and symmetries of the lowest rotationally
  excited states of $^{16}$O$_2$-$^{14}$N$_2$. Allowed states are given in
  boldface.} 
\label{table6}
\begin{tabular}{c c | c c | c c }
\hline
\multicolumn{2}{c}{J=1} & \multicolumn{2}{c}{J=2} & \multicolumn{2}{c}{J=3}\\
\hline
cm$^{-1}$ & Sym & cm$^{-1}$ & Sym & cm$^{-1}$ & Sym \\ 
\hline
 {\bf -83.120} & ${\bf B_1'}$ & {\bf -82.804} & ${\bf B_2'}$ &-82.332 &
 $B_1''$ \\ 
 {\bf -81.492} & ${\bf A_2'}$ & {\bf -81.185} & ${\bf A_2'}$ & {\bf -80.724} &
$ {\bf A_2^{'} }$ \\
 {\bf -81.485} & ${\bf A_1'}$ & {\bf -81.165} & ${\bf A_1'}$ & {\bf -80.685} &
${\bf A_1^{'} } $ \\ 
 -80.366 & $A_2''$ & -80.063 & $A_1''$ &  -79.608 & $A_2''$ \\ 
 -80.358 & $A_1''$ & -80.038 & $A_2''$ &  -79.558 & $A_1''$ \\ 
 -78.952 & $B_2''$ & -78.638 & $B_1''$ & {\bf -78.168} & ${\bf B_2'} $ \\ 
 -78.211 & $B_1''$ & -78.516 & $B_2''$ & {\bf -78.057} & ${\bf B_1' }$ \\ 
 -75.624 & $A_2''$ & -78.515 & $B_1''$ &  {\bf -78.054} & ${\bf B_2' } $ \\ 
 -74.240 & $B_1''$ & -77.900 & $B_2''$ & {\bf -77.433} & ${\bf B_1' }$ \\ 
 -74.235 & $B_2''$ & {\bf -76.560} & ${\bf B_2'}$ & -76.095 & $B_1''$ \\ 
         &         & {\bf -76.560} & ${\bf B_1'}$ &  -76.095 & $B_2''$ \\ 
         &         & -75.321 & $A_1''$ &  -74.867 & $A_2''$ \\
\hline          
\end{tabular}
\end{table}

In Table \ref{table5} we report the energies of the lowest 
bound states for total angular momentum $J=0$, together with their behavior
under the operations of the symmetry group. Note that only the states shown in
boldface are allowed. It can be noticed that there is a
rather large density of states or, in other words, the vibrational motions
involved have quite low frequencies. Analysis of the corresponding wave
functions confirms the characteristic floppiness of this dimer. 
They have been obtained following the method of Thornley and
Hutson \cite{Thornley}, and setting the SF polar angles to ${\bf
  \hat{R}}$=$(\theta_R,\phi_R)$=(0,0), 
${\bf \hat{r}}_a$=$(\theta_a,\phi_a)$=$(\theta_a,\phi/2)$ and 
${\bf \hat{r}}_b$=$(\theta_b,\phi_b)$=$(\theta_b,-\phi/2)$. 
For the three lowest states ($n=0,1,2$), wave functions multiplied by $R$ ($R
\Psi$)  are shown in Fig. \ref{fig7}. They all have their largest amplitude for 
$\theta_a$=$\theta_b$=$\pi/2$ and, for these reasons, these wavefunctions are
depicted as functions of the intermolecular distance, $R$, and the torsional
angle, $\phi$, having fixed $\theta_a$ and $\theta_b$ at $\pi/2$. The ground
state ($n$=0, Fig. \ref{fig7}(a)) is very delocalized with respect to the
torsional motion, with a slight propensity for a {\sf X} configuration. The first
excited state ($n$=1, Fig.\ref{fig7}(b)) exhibits a node for $\phi$=0 and
has its maximum probability density at the {\sf X} geometry. The second
excited state also corresponds to an excitation of the torsional motion, in
this case, the node is at $\phi$=$\pi/2$ and its geometry matches the {\sf H}
orientation. 

The relatively large stability of the T-shape orientations in the O$_2$-N$_2$
PES (Fig. \ref{fig3}) is 
revealed in the structure of the higher excited states. Indeed, the third
and fourth excited states have T-shape geometries. Wave functions for these
states are presented in Fig. \ref{fig8}. The wave function of the $n$=3 state is
depicted as a function of $R$ and $\theta_b$, where the $\theta_a$ and $\phi$ angles
have been fixed to $\pi/2$ and 0, respectively  (Fig. \ref{fig8}(a)). We have
chosen to fix $\theta_a$ to $\pi/2$ because a maximum of the probability density
is found precisely at this orientation. On the other hand, we have found that
the amplitude is quite isotropic with respect to the torsional angle
and have just chosen a representative planar configuration ($\phi=0$). It can
be seen that this function displays a node at
$\theta_b$=$\pi/2$ and a maximum amplitude for $\theta_b$=0, corresponding
to a {\sf T$_a$} configuration. Nevertheless, this angular mode displays a
wide amplitude motion, since the amplitude is also significant for
displacements out of the {\sf T$_a$} geometry. The $n$=4 state has, on the
other hand, a {\sf T$_b$} geometry. The 
shape of its wave function as a function of $R$ and $\theta_a$ is given in
Fig. \ref{fig8}(b), having fixed $\theta_b$ and $\phi$ to $\pi/2$ and 0,
respectively. As the previos level, the $n$=4 state is also quite floppy with
respect to motions of all the intermolecular degrees of freedom.

Finally, a list of some rotationally excited states ($J$= 1, 2, and 3) is
reported in Table \ref{table6}. It is worthwhile noting that the lowest levels
of $J$= 1 and 2 are allowed states and their energies are lower than the
lowest allowed states of $J=0$ (Table \ref{table5}). This feature was already
discussed by Aquilanti {\em et al} in their study of  O$_2$-N$_2$ based in the
Perugia-PES\cite{Aquilanti:03}.

Unfortunately, we cannot test present results against experiments as we were
not able to find any publication on the spectroscopy of O$_2$-N$_2$. Aquilanti
et al. \cite{Aquilanti:03} reported bound states calculated with their
experimentally derived O$_2$-N$_2$ PES\cite{Aquilanti:99bis}. They reported
$J=0$ states build up using odd and even values of $j$ for O$_2$ and N$_2$,
respectively. The lowest 
energy levels are -98.1, -88.4 and -83.7 cm$^{-1}$, which should be compared
with the corresponding present levels, -73.9, -67.2 and -67.1  cm$^{-1}$. As
can be seen, differences between the {\em ab initio} PES and the
Perugia-PES, discussed previously, become crucial for a
quantitative prediction of the bound spectrum of this complex.

Present results can be also compared with the vibrational structure of the
related dimers O$_2$-O$_2$ and N$_2$-N$_2$. First, O$_2$-N$_2$ bound states
are quite different to those of O$_2$-O$_2$ in the singlet and triplet spin
multiplicities, as, due to exchange interactions involving the two open-shell
molecules, they exhibit a preference for the {\sf H} geometry and are much more
rigid\cite{Carmona-Novillo:12}. However, the lowest vibrational states of
O$_2$-N$_2$ are impressively similar to accurate calculations of  O$_2$-O$_2$
in the quintet multiplicity \cite{max-jcp:08}.
 In addition, the shape of the associated
wave functions is very alike.   Finally, comparing with calculations based on
an experimentally derived PES for N$_2$-N$_2$\cite{n2our}, it is found that energies of
the lowest bound states are rather similar: the three lowest energies are
-79.9, -77.4 and -71.1. However, it was found that the ground state has a
T-shaped geometry (in accordance with the features of the PES), whereas the
first state in  O$_2$-N$_2$ exhibiting such a geometry is the third
vibrationally excited state.

\section{ Concluding remarks}
\label{conclu}

In the present paper state-of-the-art {\em ab initio} techniques have 
been applied to compute the ground state potential energy surface for the 
interaction of rigid N$_2$ and O$_2$ molecules in their ground electronic 
and vibrational states.
The main results of this paper can be summarized as follows.

\begin{enumerate}

\item
The potential energy surface for the O$_2(^3\Sigma^-_g)$ + N$_2(^1\Sigma^+_g)$ 
complex has a global minimum of 14.63 meV at $R=6.52$ bohr for the {\sf X}
($D_{2d}$) geometry, and local minima of 14.05 meV at $R=7.44$ bohr and
12.58 meV at $R=7.52$ bohr for the {\sf T$_a$} and {\sf T$_b$} geometries,
respectively. 

\item
Symmetry-adapted perturbation theory utilizing the DFT description of the isolated
monomers was shown to work very well in comparison with the spin-restricted coupled
cluster method with single, double, and noniterative triple excitations.
\item
The accuracy of the computed potential was checked by comparison with the empirical
potential from the Perugia group fitted to the measured integral cross sections.
In general, relatively good agreement between the {\em ab initio} and empirical surfaces is
found, the deviations ranging from 3\% for the {\sf H} geometry to 43\% for the
weakly bound {\sf L} geometry. The present potential is more anisotropic than
the experimentally derived potential.
\item
The {\em ab initio} potential satisfactorily reproduces the measured integral cross
sections. Both the glory oscillations as a function of the beam intensity and the
absolute cross sections are in a (semi)quantitative agreement with the experimental
data.

\item
The {\em ab initio} potential also works well for the second virial
coefficient of the O$_2$-N$_2$ mixture. Except for the two lowest 
temperatures, the present results are well within the experimental 
error bars leading to the best agreement to date for
temperatures higher than the Boyle temperature. Second virial coefficients
are also computed at temperatures where measured values are unavailable and,
considering the accuracy of the present PES, they can be
safely used to guide extrapolations at high temperature.

\item
The analysis of the bound states of the complex reveals that the complex is floppy,
and the nearly free rotors description is adequate. Unfortunately, no experimental
data are available for comparison.
\end{enumerate}

The potential energy surface reported in this work can be employed in future
investigations such as studies of collision induced absorption or rotationally
energy transfer in collisions.

\section*{Acknowledgments}

We thank financial support by Spanish grant FIS2010-22064-C02-02. 
RM thanks the Foundation for Polish Science for support within 
the MISTRZ programme. 
Allocation of computing time by CESGA (Spain) and the COST-CMTS Action CM1002
``Convergent Distributed Enviroment for Computational Spectroscopy (CODECS)''
are also acknowledged.


\begin{thebibliography}{61}
\expandafter\ifx\csname natexlab\endcsname\relax\def\natexlab#1{#1}\fi
\expandafter\ifx\csname bibnamefont\endcsname\relax
  \def\bibnamefont#1{#1}\fi
\expandafter\ifx\csname bibfnamefont\endcsname\relax
  \def\bibfnamefont#1{#1}\fi
\expandafter\ifx\csname citenamefont\endcsname\relax
  \def\citenamefont#1{#1}\fi
\expandafter\ifx\csname url\endcsname\relax
  \def\url#1{\texttt{#1}}\fi
\expandafter\ifx\csname urlprefix\endcsname\relax\def\urlprefix{URL }\fi
\providecommand{\bibinfo}[2]{#2}
\providecommand{\eprint}[2][]{\url{#2}}

\bibitem[{\citenamefont{Lewis}(1924)}]{Lewis:24}
\bibinfo{author}{\bibfnamefont{G.~N.} \bibnamefont{Lewis}},
  \bibinfo{journal}{J. Am. Chem. Soc.} \textbf{\bibinfo{volume}{46}},
  \bibinfo{pages}{2027} (\bibinfo{year}{1924}).

\bibitem[{\citenamefont{Crawford et~al.}(1949)\citenamefont{Crawford, Welsh,
  and Locke}}]{Welsh-dimer:49}
\bibinfo{author}{\bibfnamefont{M.~F.} \bibnamefont{Crawford}},
  \bibinfo{author}{\bibfnamefont{H.~L.} \bibnamefont{Welsh}}, \bibnamefont{and}
  \bibinfo{author}{\bibfnamefont{J.~L.} \bibnamefont{Locke}},
  \bibinfo{journal}{Phys. Rev.} \textbf{\bibinfo{volume}{75}},
  \bibinfo{pages}{1607} (\bibinfo{year}{1949}).

\bibitem[{\citenamefont{Frommhold}(1994)}]{cia-frommhold}
\bibinfo{author}{\bibfnamefont{L.}~\bibnamefont{Frommhold}},
  \emph{\bibinfo{title}{{\it Collision-induced Absorption in Gases}}}
  (\bibinfo{publisher}{Cambridge Monographs on Atomic, Molecular and Chemical
  Physics, Cambridge University Press}, \bibinfo{year}{1994}), ISBN
  \bibinfo{isbn}{9780521393454}.

\bibitem[{\citenamefont{Abel and Frommhold}(2013)}]{frommhold-13-cia-astro}
\bibinfo{author}{\bibfnamefont{M.}~\bibnamefont{Abel}} \bibnamefont{and}
  \bibinfo{author}{\bibfnamefont{L.}~\bibnamefont{Frommhold}},
  \bibinfo{journal}{Can. J. Phys.} \textbf{\bibinfo{volume}{91}},
  \bibinfo{pages}{857} (\bibinfo{year}{2013}).

\bibitem[{\citenamefont{Klemperer and
  Vaida}(2006)}]{klemperer-06-molclus-atmos}
\bibinfo{author}{\bibfnamefont{W.}~\bibnamefont{Klemperer}} \bibnamefont{and}
  \bibinfo{author}{\bibfnamefont{V.}~\bibnamefont{Vaida}},
  \bibinfo{journal}{Proc. Nat. Ac. Sci.} \textbf{\bibinfo{volume}{103}},
  \bibinfo{pages}{10584} (\bibinfo{year}{2006}).

\bibitem[{\citenamefont{Bartolomei
  et~al.}(2008{\natexlab{a}})\citenamefont{Bartolomei, Carmona-Novillo,
  Hern{\'a}ndez, Campos-Mart{\'\i}nez, and
  Hern{\'a}ndez-Lamoneda}}]{max-jcp:08}
\bibinfo{author}{\bibfnamefont{M.}~\bibnamefont{Bartolomei}},
  \bibinfo{author}{\bibfnamefont{E.}~\bibnamefont{Carmona-Novillo}},
  \bibinfo{author}{\bibfnamefont{M.~I.} \bibnamefont{Hern{\'a}ndez}},
  \bibinfo{author}{\bibfnamefont{J.}~\bibnamefont{Campos-Mart{\'\i}nez}},
  \bibnamefont{and}
  \bibinfo{author}{\bibfnamefont{R.}~\bibnamefont{Hern{\'a}ndez-Lamoneda}},
  \bibinfo{journal}{J. Chem. Phys.} \textbf{\bibinfo{volume}{128}},
  \bibinfo{pages}{214304} (\bibinfo{year}{2008}{\natexlab{a}}).

\bibitem[{\citenamefont{Bartolomei
  et~al.}(2008{\natexlab{b}})\citenamefont{Bartolomei, Hern{\'a}ndez,
  Campos-Mart{\'\i}nez, Carmona-Novillo, and
  Hern{\'a}ndez-Lamoneda}}]{maxpccp:08}
\bibinfo{author}{\bibfnamefont{M.}~\bibnamefont{Bartolomei}},
  \bibinfo{author}{\bibfnamefont{M.~I.} \bibnamefont{Hern{\'a}ndez}},
  \bibinfo{author}{\bibfnamefont{J.}~\bibnamefont{Campos-Mart{\'\i}nez}},
  \bibinfo{author}{\bibfnamefont{E.}~\bibnamefont{Carmona-Novillo}},
  \bibnamefont{and}
  \bibinfo{author}{\bibfnamefont{R.}~\bibnamefont{Hern{\'a}ndez-Lamoneda}},
  \bibinfo{journal}{Phys. Chem. Chem. Phys} \textbf{\bibinfo{volume}{10}},
  \bibinfo{pages}{5374} (\bibinfo{year}{2008}{\natexlab{b}}).

\bibitem[{\citenamefont{Bartolomei et~al.}(2010)\citenamefont{Bartolomei,
  Carmona-Novillo, Campos-Mart{\'\i}nez, Hern{\'a}ndez, and
  Hern{\'a}ndez-Lamoneda}}]{O2O2pes:10}
\bibinfo{author}{\bibfnamefont{M.}~\bibnamefont{Bartolomei}},
  \bibinfo{author}{\bibfnamefont{E.}~\bibnamefont{Carmona-Novillo}},
  \bibinfo{author}{\bibfnamefont{J.}~\bibnamefont{Campos-Mart{\'\i}nez}},
  \bibinfo{author}{\bibfnamefont{M.~I.} \bibnamefont{Hern{\'a}ndez}},
  \bibnamefont{and}
  \bibinfo{author}{\bibfnamefont{R.}~\bibnamefont{Hern{\'a}ndez-Lamoneda}},
  \bibinfo{journal}{J. Chem. Phys.} \textbf{\bibinfo{volume}{133}},
  \bibinfo{pages}{12431} (\bibinfo{year}{2010}).

\bibitem[{\citenamefont{Bartolomei et~al.}(2011)\citenamefont{Bartolomei,
  Carmona-Novillo, Hern{\'a}ndez, Campos-Mart{\'\i}nez, and
  Hern{\'a}ndez-Lamoneda}}]{max-lr:10}
\bibinfo{author}{\bibfnamefont{M.}~\bibnamefont{Bartolomei}},
  \bibinfo{author}{\bibfnamefont{E.}~\bibnamefont{Carmona-Novillo}},
  \bibinfo{author}{\bibfnamefont{M.~I.} \bibnamefont{Hern{\'a}ndez}},
  \bibinfo{author}{\bibfnamefont{J.}~\bibnamefont{Campos-Mart{\'\i}nez}},
  \bibnamefont{and}
  \bibinfo{author}{\bibfnamefont{R.}~\bibnamefont{Hern{\'a}ndez-Lamoneda}},
  \bibinfo{journal}{J. Comp. Chem.} \textbf{\bibinfo{volume}{32}},
  \bibinfo{pages}{279} (\bibinfo{year}{2011}).

\bibitem[{\citenamefont{Gomez et~al.}(2007)\citenamefont{Gomez,
  Bussery-Honvault, Cauchy, Bartolomei, Cappelletti, and Pirani}}]{Bussery:07}
\bibinfo{author}{\bibfnamefont{L.}~\bibnamefont{Gomez}},
  \bibinfo{author}{\bibfnamefont{B.}~\bibnamefont{Bussery-Honvault}},
  \bibinfo{author}{\bibfnamefont{T.}~\bibnamefont{Cauchy}},
  \bibinfo{author}{\bibfnamefont{M.}~\bibnamefont{Bartolomei}},
  \bibinfo{author}{\bibfnamefont{D.}~\bibnamefont{Cappelletti}},
  \bibnamefont{and} \bibinfo{author}{\bibfnamefont{F.}~\bibnamefont{Pirani}},
  \bibinfo{journal}{Chem. Phys. Lett.} \textbf{\bibinfo{volume}{445}},
  \bibinfo{pages}{99} (\bibinfo{year}{2007}).

\bibitem[{\citenamefont{Jeziorski et~al.}(1994)\citenamefont{Jeziorski,
  Moszynski, and Szalewicz}}]{sapt:94}
\bibinfo{author}{\bibfnamefont{B.}~\bibnamefont{Jeziorski}},
  \bibinfo{author}{\bibfnamefont{R.}~\bibnamefont{Moszynski}},
  \bibnamefont{and}
  \bibinfo{author}{\bibfnamefont{K.}~\bibnamefont{Szalewicz}},
  \bibinfo{journal}{Chem. Rev.} \textbf{\bibinfo{volume}{94}},
  \bibinfo{pages}{1887} (\bibinfo{year}{1994}).

\bibitem[{\citenamefont{Bussery-Honvault and
  Hartmann}(2014)}]{hart-buss-cia-n2:13}
\bibinfo{author}{\bibfnamefont{B.}~\bibnamefont{Bussery-Honvault}}
  \bibnamefont{and} \bibinfo{author}{\bibfnamefont{J.-M.}
  \bibnamefont{Hartmann}}, \bibinfo{journal}{J. Chem. Phys.}
  \textbf{\bibinfo{volume}{140}}, \bibinfo{pages}{054309}
  (\bibinfo{year}{2014}), \bibinfo{note}{054309:1-054309:6}.

\bibitem[{\citenamefont{Aquilanti et~al.}(2003)\citenamefont{Aquilanti,
  Bartolomei, Carmona-Novillo, and Pirani}}]{Aquilanti:03}
\bibinfo{author}{\bibfnamefont{V.}~\bibnamefont{Aquilanti}},
  \bibinfo{author}{\bibfnamefont{M.}~\bibnamefont{Bartolomei}},
  \bibinfo{author}{\bibfnamefont{E.}~\bibnamefont{Carmona-Novillo}},
  \bibnamefont{and} \bibinfo{author}{\bibfnamefont{F.}~\bibnamefont{Pirani}},
  \bibinfo{journal}{J. Chem. Phys.} \textbf{\bibinfo{volume}{118}},
  \bibinfo{pages}{2214} (\bibinfo{year}{2003}).

\bibitem[{\citenamefont{Zuchowski et~al.}(2008)\citenamefont{Zuchowski,
  Podeszwa, Moszynski, Jeziorski, and Szalewicz}}]{SAPTDFT:08}
\bibinfo{author}{\bibfnamefont{P.~S.} \bibnamefont{Zuchowski}},
  \bibinfo{author}{\bibfnamefont{R.}~\bibnamefont{Podeszwa}},
  \bibinfo{author}{\bibfnamefont{R.}~\bibnamefont{Moszynski}},
  \bibinfo{author}{\bibfnamefont{B.}~\bibnamefont{Jeziorski}},
  \bibnamefont{and}
  \bibinfo{author}{\bibfnamefont{K.}~\bibnamefont{Szalewicz}},
  \bibinfo{journal}{J. Chem. Phys.} \textbf{\bibinfo{volume}{129}},
  \bibinfo{pages}{084101} (\bibinfo{year}{2008}), ISSN
  \bibinfo{issn}{0021-9606}.

\bibitem[{\citenamefont{van~der Avoird and Brocks}(1987)}]{vdABrocks:87}
\bibinfo{author}{\bibfnamefont{A.}~\bibnamefont{van~der Avoird}}
  \bibnamefont{and} \bibinfo{author}{\bibfnamefont{G.}~\bibnamefont{Brocks}},
  \bibinfo{journal}{J. Chem. Phys.} \textbf{\bibinfo{volume}{87}},
  \bibinfo{pages}{5346} (\bibinfo{year}{1987}).

\bibitem[{\citenamefont{Kendall et~al.}(1992)\citenamefont{Kendall, Dunning,
  and Harrison}}]{Dunning}
\bibinfo{author}{\bibfnamefont{R.~A.} \bibnamefont{Kendall}},
  \bibinfo{author}{\bibfnamefont{T.~H.} \bibnamefont{Dunning}},
  \bibnamefont{and} \bibinfo{author}{\bibfnamefont{R.~J.}
  \bibnamefont{Harrison}}, \bibinfo{journal}{J. Chem. Phys.}
  \textbf{\bibinfo{volume}{96}}, \bibinfo{pages}{6796} (\bibinfo{year}{1992}).

\bibitem[{\citenamefont{Tao and Pan}(1992)}]{tao:92}
\bibinfo{author}{\bibfnamefont{F.~M.} \bibnamefont{Tao}} \bibnamefont{and}
  \bibinfo{author}{\bibfnamefont{Y.~K.} \bibnamefont{Pan}},
  \bibinfo{journal}{J. Chem. Phys.} \textbf{\bibinfo{volume}{97}},
  \bibinfo{pages}{4989} (\bibinfo{year}{1992}).

\bibitem[{\citenamefont{Adamo and Barone}(1999)}]{PBE0}
\bibinfo{author}{\bibfnamefont{C.}~\bibnamefont{Adamo}} \bibnamefont{and}
  \bibinfo{author}{\bibfnamefont{V.}~\bibnamefont{Barone}},
  \bibinfo{journal}{J. Chem. Phys.} \textbf{\bibinfo{volume}{110}},
  \bibinfo{pages}{6158} (\bibinfo{year}{1999}).

\bibitem[{Dal(2005)}]{Dalton}
\emph{\bibinfo{title}{Dalton, a molecular electronic structure program, release
  2.0}} (\bibinfo{year}{2005}), \bibinfo{note}{see
  http://www.kjemi.uio.no/software/dalton/dalton.html}.

\bibitem[{\citenamefont{Fermi and Amaldi}(1934)}]{Fermi:34}
\bibinfo{author}{\bibfnamefont{E.}~\bibnamefont{Fermi}} \bibnamefont{and}
  \bibinfo{author}{\bibfnamefont{G.}~\bibnamefont{Amaldi}},
  \bibinfo{journal}{Memorie della Classe di Scienze Fisiche, Matematiche e
  Naturali. R. Accademia D'Italia} \textbf{\bibinfo{volume}{6}},
  \bibinfo{pages}{117} (\bibinfo{year}{1934}).

\bibitem[{\citenamefont{Tozer and Handy}(1998)}]{Handy:98}
\bibinfo{author}{\bibfnamefont{D.~J.} \bibnamefont{Tozer}} \bibnamefont{and}
  \bibinfo{author}{\bibfnamefont{N.~C.} \bibnamefont{Handy}},
  \bibinfo{journal}{J. Chem. Phys.} \textbf{\bibinfo{volume}{109}},
  \bibinfo{pages}{10180} (\bibinfo{year}{1998}).

\bibitem[{\citenamefont{{\.Z}uchowski}(2008)}]{zuchowski:08}
\bibinfo{author}{\bibfnamefont{P.~S.} \bibnamefont{{\.Z}uchowski}},
  \bibinfo{journal}{Chem. Phys. Lett.} \textbf{\bibinfo{volume}{450}},
  \bibinfo{pages}{203} (\bibinfo{year}{2008}).

\bibitem[{NIS()}]{NIST}
\bibinfo{note}{NIST Chemistry WebBook, NIST Standard Reference Database Number
  69, ed. P. J. Linstrom and W. G. Mallard, 2007, http://WebBook.nist.gov.}

\bibitem[{\citenamefont{Bukowski et~al.}(2008)\citenamefont{Bukowski, Cencek,
  Jankowski, Jeziorska, Jeziorski, Kucharski, Lotrich, Misquitta, Moszynski,
  Patkowski et~al.}}]{SAPT-code}
\bibinfo{author}{\bibfnamefont{R.}~\bibnamefont{Bukowski}},
  \bibinfo{author}{\bibfnamefont{W.}~\bibnamefont{Cencek}},
  \bibinfo{author}{\bibfnamefont{P.}~\bibnamefont{Jankowski}},
  \bibinfo{author}{\bibfnamefont{M.}~\bibnamefont{Jeziorska}},
  \bibinfo{author}{\bibfnamefont{B.}~\bibnamefont{Jeziorski}},
  \bibinfo{author}{\bibfnamefont{S.~A.} \bibnamefont{Kucharski}},
  \bibinfo{author}{\bibfnamefont{V.~F.} \bibnamefont{Lotrich}},
  \bibinfo{author}{\bibfnamefont{A.~J.} \bibnamefont{Misquitta}},
  \bibinfo{author}{\bibfnamefont{R.}~\bibnamefont{Moszynski}},
  \bibinfo{author}{\bibfnamefont{K.}~\bibnamefont{Patkowski}},
  \bibnamefont{et~al.}, \emph{\bibinfo{title}{Sapt2008, version 2008.2, an ab
  initio program for many-body-symmetry-adapted perturbation theory of
  intermolecular interaction energies}} (\bibinfo{year}{2008}),
  \bibinfo{note}{see http://www.physics.udel.edu/~szalewic/SAPT/index.html}.

\bibitem[{\citenamefont{Boys and Bernardi}(1970)}]{Boys:70}
\bibinfo{author}{\bibfnamefont{S.}~\bibnamefont{Boys}} \bibnamefont{and}
  \bibinfo{author}{\bibfnamefont{F.}~\bibnamefont{Bernardi}},
  \bibinfo{journal}{Mol. Phys.} \textbf{\bibinfo{volume}{19}},
  \bibinfo{pages}{553} (\bibinfo{year}{1970}).

\bibitem[{\citenamefont{van Lenthe et~al.}(1987)\citenamefont{van Lenthe, van
  Duijneveldt-van~de Rijdt, and van Duijneveldt}}]{Lenthe:87}
\bibinfo{author}{\bibfnamefont{J.~H.} \bibnamefont{van Lenthe}},
  \bibinfo{author}{\bibfnamefont{J.~G.~C.~M.} \bibnamefont{van
  Duijneveldt-van~de Rijdt}}, \bibnamefont{and}
  \bibinfo{author}{\bibfnamefont{F.}~\bibnamefont{van Duijneveldt}},
  \bibinfo{journal}{Adv. Chem. Phys.} \textbf{\bibinfo{volume}{69}},
  \bibinfo{pages}{521} (\bibinfo{year}{1987}).

\bibitem[{\citenamefont{Werner et~al.}(2006)\citenamefont{Werner, Knowles,
  Lindh, Manby, {Sch\"{u}tz}, Celani, Korona, Rauhut, Amos, Bernhardsson
  et~al.}}]{MOLPRO}
\bibinfo{author}{\bibfnamefont{H.-J.} \bibnamefont{Werner}},
  \bibinfo{author}{\bibfnamefont{P.~J.} \bibnamefont{Knowles}},
  \bibinfo{author}{\bibfnamefont{R.}~\bibnamefont{Lindh}},
  \bibinfo{author}{\bibfnamefont{F.~R.} \bibnamefont{Manby}},
  \bibinfo{author}{\bibfnamefont{M.}~\bibnamefont{{Sch\"{u}tz}}},
  \bibinfo{author}{\bibfnamefont{P.}~\bibnamefont{Celani}},
  \bibinfo{author}{\bibfnamefont{T.}~\bibnamefont{Korona}},
  \bibinfo{author}{\bibfnamefont{G.}~\bibnamefont{Rauhut}},
  \bibinfo{author}{\bibfnamefont{R.~D.} \bibnamefont{Amos}},
  \bibinfo{author}{\bibfnamefont{A.}~\bibnamefont{Bernhardsson}},
  \bibnamefont{et~al.}, \emph{\bibinfo{title}{Molpro, version2006.1, a package
  of ab initio programs}} (\bibinfo{year}{2006}),
  \bibinfo{note}{seehttp://www.molpro.net}.

\bibitem[{\citenamefont{Wormer and van~der Avoird}(1984)}]{Wormer:84}
\bibinfo{author}{\bibfnamefont{P.~E.~S.} \bibnamefont{Wormer}}
  \bibnamefont{and} \bibinfo{author}{\bibfnamefont{A.}~\bibnamefont{van~der
  Avoird}}, \bibinfo{journal}{J. Chem. Phys.} \textbf{\bibinfo{volume}{81}},
  \bibinfo{pages}{1929} (\bibinfo{year}{1984}).

\bibitem[{\citenamefont{Abramovitz and Stegun}(1972)}]{abramovitz}
\bibinfo{author}{\bibfnamefont{M.}~\bibnamefont{Abramovitz}} \bibnamefont{and}
  \bibinfo{author}{\bibfnamefont{I.~A.} \bibnamefont{Stegun}},
  \emph{\bibinfo{title}{\em Handbook of Mathematical Functions}}
  (\bibinfo{publisher}{Dover, New York}, \bibinfo{year}{1972}).

\bibitem[{\citenamefont{Cwiok et~al.}(1992)\citenamefont{Cwiok, Jeziorski,
  Kolos, Moszynski, Rychlewski, and Szalewicz}}]{Cwiok:92}
\bibinfo{author}{\bibfnamefont{T.}~\bibnamefont{Cwiok}},
  \bibinfo{author}{\bibfnamefont{B.}~\bibnamefont{Jeziorski}},
  \bibinfo{author}{\bibfnamefont{W.}~\bibnamefont{Kolos}},
  \bibinfo{author}{\bibfnamefont{R.}~\bibnamefont{Moszynski}},
  \bibinfo{author}{\bibfnamefont{J.}~\bibnamefont{Rychlewski}},
  \bibnamefont{and}
  \bibinfo{author}{\bibfnamefont{K.}~\bibnamefont{Szalewicz}},
  \bibinfo{journal}{Chem. Phys. Lett.} \textbf{\bibinfo{volume}{195}},
  \bibinfo{pages}{67} (\bibinfo{year}{1992}).

\bibitem[{\citenamefont{van~der Avoird et~al.}(1980)\citenamefont{van~der
  Avoird, Wormer, Mulder, and Berns}}]{Avoird:80}
\bibinfo{author}{\bibfnamefont{A.}~\bibnamefont{van~der Avoird}},
  \bibinfo{author}{\bibfnamefont{P.~E.~S.} \bibnamefont{Wormer}},
  \bibinfo{author}{\bibfnamefont{F.}~\bibnamefont{Mulder}}, \bibnamefont{and}
  \bibinfo{author}{\bibfnamefont{R.~M.} \bibnamefont{Berns}}, in
  \emph{\bibinfo{booktitle}{\it {Topics in Current Chemistry}}}
  (\bibinfo{publisher}{Springer, Berlin, Vol.93, Chapter 1},
  \bibinfo{year}{1980}), pp. \bibinfo{pages}{1--51}.

\bibitem[{\citenamefont{Heijmen et~al.}(1996)\citenamefont{Heijmen, Moszynski,
  Wormer, and van~der Avoird}}]{Heijmen:96}
\bibinfo{author}{\bibfnamefont{T.~G.~A.} \bibnamefont{Heijmen}},
  \bibinfo{author}{\bibfnamefont{R.}~\bibnamefont{Moszynski}},
  \bibinfo{author}{\bibfnamefont{P.~E.~S.} \bibnamefont{Wormer}},
  \bibnamefont{and} \bibinfo{author}{\bibfnamefont{A.}~\bibnamefont{van~der
  Avoird}}, \bibinfo{journal}{Mol. Phys.} \textbf{\bibinfo{volume}{89}},
  \bibinfo{pages}{81} (\bibinfo{year}{1996}).

\bibitem[{\citenamefont{Brunetti et~al.}(1981)\citenamefont{Brunetti, Liuti,
  Luzzatti, Pirani, and Vecchiocattivi}}]{Brunetti:81}
\bibinfo{author}{\bibfnamefont{B.}~\bibnamefont{Brunetti}},
  \bibinfo{author}{\bibfnamefont{G.}~\bibnamefont{Liuti}},
  \bibinfo{author}{\bibfnamefont{E.}~\bibnamefont{Luzzatti}},
  \bibinfo{author}{\bibfnamefont{F.}~\bibnamefont{Pirani}}, \bibnamefont{and}
  \bibinfo{author}{\bibfnamefont{F.}~\bibnamefont{Vecchiocattivi}},
  \bibinfo{journal}{J. Chem. Phys.} \textbf{\bibinfo{volume}{74}},
  \bibinfo{pages}{6734} (\bibinfo{year}{1981}).

\bibitem[{\citenamefont{Cappelletti et~al.}(2008)\citenamefont{Cappelletti,
  Pirani, Bussery-Honvault, G{\'o}mez, and Bartolomei}}]{cappelletti:08}
\bibinfo{author}{\bibfnamefont{D.}~\bibnamefont{Cappelletti}},
  \bibinfo{author}{\bibfnamefont{F.}~\bibnamefont{Pirani}},
  \bibinfo{author}{\bibfnamefont{B.}~\bibnamefont{Bussery-Honvault}},
  \bibinfo{author}{\bibfnamefont{L.}~\bibnamefont{G{\'o}mez}},
  \bibnamefont{and}
  \bibinfo{author}{\bibfnamefont{M.}~\bibnamefont{Bartolomei}},
  \bibinfo{journal}{Phys. Chem. Chem. Phys} \textbf{\bibinfo{volume}{10}},
  \bibinfo{pages}{4281} (\bibinfo{year}{2008}).

\bibitem[{\citenamefont{Bartolomei
  et~al.}(2008{\natexlab{c}})\citenamefont{Bartolomei, Cappelletti, de~Petris,
  Texidor, Pirani, Rosi, and Vecchiocattivi}}]{pccpnon2:08}
\bibinfo{author}{\bibfnamefont{M.}~\bibnamefont{Bartolomei}},
  \bibinfo{author}{\bibfnamefont{D.}~\bibnamefont{Cappelletti}},
  \bibinfo{author}{\bibfnamefont{G.}~\bibnamefont{de~Petris}},
  \bibinfo{author}{\bibfnamefont{M.~M.} \bibnamefont{Texidor}},
  \bibinfo{author}{\bibfnamefont{F.}~\bibnamefont{Pirani}},
  \bibinfo{author}{\bibfnamefont{M.}~\bibnamefont{Rosi}}, \bibnamefont{and}
  \bibinfo{author}{\bibfnamefont{F.}~\bibnamefont{Vecchiocattivi}},
  \bibinfo{journal}{Phys. Chem. Chem. Phys} \textbf{\bibinfo{volume}{10}},
  \bibinfo{pages}{5993} (\bibinfo{year}{2008}{\natexlab{c}}).

\bibitem[{\citenamefont{Thibault et~al.}(2009)\citenamefont{Thibault,
  Cappelletti, Pirani, and Bartolomei}}]{jpcac2h2:09}
\bibinfo{author}{\bibfnamefont{F.}~\bibnamefont{Thibault}},
  \bibinfo{author}{\bibfnamefont{D.}~\bibnamefont{Cappelletti}},
  \bibinfo{author}{\bibfnamefont{F.}~\bibnamefont{Pirani}}, \bibnamefont{and}
  \bibinfo{author}{\bibfnamefont{M.}~\bibnamefont{Bartolomei}},
  \bibinfo{journal}{J. Phys. Chem. A} \textbf{\bibinfo{volume}{113}},
  \bibinfo{pages}{14867} (\bibinfo{year}{2009}).

\bibitem[{\citenamefont{Cappelletti et~al.}(2002)\citenamefont{Cappelletti,
  Bartolomei, Pirani, and Aquilanti}}]{jpc-benzenehot:02}
\bibinfo{author}{\bibfnamefont{D.}~\bibnamefont{Cappelletti}},
  \bibinfo{author}{\bibfnamefont{M.}~\bibnamefont{Bartolomei}},
  \bibinfo{author}{\bibfnamefont{F.}~\bibnamefont{Pirani}}, \bibnamefont{and}
  \bibinfo{author}{\bibfnamefont{V.}~\bibnamefont{Aquilanti}},
  \bibinfo{journal}{J. Phys. Chem. A} \textbf{\bibinfo{volume}{106}},
  \bibinfo{pages}{10764} (\bibinfo{year}{2002}).

\bibitem[{\citenamefont{Schiff}(1956)}]{schiff56}
\bibinfo{author}{\bibfnamefont{L.~I.} \bibnamefont{Schiff}},
  \bibinfo{journal}{Phys. Rev.} \textbf{\bibinfo{volume}{103}},
  \bibinfo{pages}{443} (\bibinfo{year}{1956}).

\bibitem[{\citenamefont{Pirani and Vecchiocattivi}(1982)}]{Pirani:82}
\bibinfo{author}{\bibfnamefont{F.}~\bibnamefont{Pirani}} \bibnamefont{and}
  \bibinfo{author}{\bibfnamefont{F.}~\bibnamefont{Vecchiocattivi}},
  \bibinfo{journal}{Mol. Phys.} \textbf{\bibinfo{volume}{45}},
  \bibinfo{pages}{1003} (\bibinfo{year}{1982}).

\bibitem[{\citenamefont{Aquilanti et~al.}(1998)\citenamefont{Aquilanti,
  Ascenzi, Cappelletti, de~Castro-V{\'{\i}tores}, and Pirani}}]{o2kr}
\bibinfo{author}{\bibfnamefont{V.}~\bibnamefont{Aquilanti}},
  \bibinfo{author}{\bibfnamefont{D.}~\bibnamefont{Ascenzi}},
  \bibinfo{author}{\bibfnamefont{D.}~\bibnamefont{Cappelletti}},
  \bibinfo{author}{\bibfnamefont{M.}~\bibnamefont{de~Castro-V{\'{\i}tores}}},
  \bibnamefont{and} \bibinfo{author}{\bibfnamefont{F.}~\bibnamefont{Pirani}},
  \bibinfo{journal}{J. Chem. Phys.} \textbf{\bibinfo{volume}{109}},
  \bibinfo{pages}{3898} (\bibinfo{year}{1998}).

\bibitem[{\citenamefont{P{\'e}rez-R{\'\i}os
  et~al.}(2009)\citenamefont{P{\'e}rez-R{\'\i}os, Bartolomei,
  Campos-Mart{\'\i}nez, Hern{\'a}ndez, and Hern{\'a}ndez-Lamoneda}}]{Jesus:09}
\bibinfo{author}{\bibfnamefont{J.}~\bibnamefont{P{\'e}rez-R{\'\i}os}},
  \bibinfo{author}{\bibfnamefont{M.}~\bibnamefont{Bartolomei}},
  \bibinfo{author}{\bibfnamefont{J.}~\bibnamefont{Campos-Mart{\'\i}nez}},
  \bibinfo{author}{\bibfnamefont{M.~I.} \bibnamefont{Hern{\'a}ndez}},
  \bibnamefont{and}
  \bibinfo{author}{\bibfnamefont{R.}~\bibnamefont{Hern{\'a}ndez-Lamoneda}},
  \bibinfo{journal}{J. Phys. Chem. A} \textbf{\bibinfo{volume}{113}},
  \bibinfo{pages}{14952} (\bibinfo{year}{2009}).

\bibitem[{\citenamefont{Bernstein and O'Brien}(1965)}]{Bernstein:65}
\bibinfo{author}{\bibfnamefont{R.~B.} \bibnamefont{Bernstein}}
  \bibnamefont{and} \bibinfo{author}{\bibfnamefont{T.~J.~P.}
  \bibnamefont{O'Brien}}, \bibinfo{journal}{Discuss. Faraday Soc.}
  \textbf{\bibinfo{volume}{40}}, \bibinfo{pages}{35} (\bibinfo{year}{1965}).

\bibitem[{\citenamefont{Bernstein and O'Brien}(1967)}]{Bernstein:67}
\bibinfo{author}{\bibfnamefont{R.~B.} \bibnamefont{Bernstein}}
  \bibnamefont{and} \bibinfo{author}{\bibfnamefont{T.~J.~P.}
  \bibnamefont{O'Brien}}, \bibinfo{journal}{J. Chem. Phys.}
  \textbf{\bibinfo{volume}{46}}, \bibinfo{pages}{1208} (\bibinfo{year}{1967}).

\bibitem[{\citenamefont{P{\'e}rez-R{\'\i}os
  et~al.}(2012)\citenamefont{P{\'e}rez-R{\'\i}os, Bartolomei,
  Campos-Mart{\'\i}nez, and Hern{\'a}ndez}}]{jpr-o2o2-cpl-12}
\bibinfo{author}{\bibfnamefont{J.}~\bibnamefont{P{\'e}rez-R{\'\i}os}},
  \bibinfo{author}{\bibfnamefont{M.}~\bibnamefont{Bartolomei}},
  \bibinfo{author}{\bibfnamefont{J.}~\bibnamefont{Campos-Mart{\'\i}nez}},
  \bibnamefont{and} \bibinfo{author}{\bibfnamefont{M.~I.}
  \bibnamefont{Hern{\'a}ndez}}, \bibinfo{journal}{Chem. Phys. Lett.}
  \textbf{\bibinfo{volume}{522}}, \bibinfo{pages}{28} (\bibinfo{year}{2012}),
  \bibinfo{note}{28-32}.

\bibitem[{\citenamefont{Martin et~al.}(1982)\citenamefont{Martin, Trengove,
  Harris, and Dunlop}}]{Dunlop:82}
\bibinfo{author}{\bibfnamefont{M.}~\bibnamefont{Martin}},
  \bibinfo{author}{\bibfnamefont{R.}~\bibnamefont{Trengove}},
  \bibinfo{author}{\bibfnamefont{K.}~\bibnamefont{Harris}}, \bibnamefont{and}
  \bibinfo{author}{\bibfnamefont{P.}~\bibnamefont{Dunlop}},
  \bibinfo{journal}{Aust. J. Chem.} \textbf{\bibinfo{volume}{35}},
  \bibinfo{pages}{1525} (\bibinfo{year}{1982}).

\bibitem[{\citenamefont{Fostiropoulos et~al.}(1988)\citenamefont{Fostiropoulos,
  Natour, Sommer, and Schramm}}]{n2o2vir}
\bibinfo{author}{\bibfnamefont{K.}~\bibnamefont{Fostiropoulos}},
  \bibinfo{author}{\bibfnamefont{G.}~\bibnamefont{Natour}},
  \bibinfo{author}{\bibfnamefont{J.}~\bibnamefont{Sommer}}, \bibnamefont{and}
  \bibinfo{author}{\bibfnamefont{B.}~\bibnamefont{Schramm}},
  \bibinfo{journal}{Ber. Bunsenges. Phys. Chem.} \textbf{\bibinfo{volume}{92}},
  \bibinfo{pages}{925} (\bibinfo{year}{1988}).

\bibitem[{\citenamefont{Pack}(1983)}]{Pack:83}
\bibinfo{author}{\bibfnamefont{R.~T.} \bibnamefont{Pack}}, \bibinfo{journal}{J.
  Chem. Phys.} \textbf{\bibinfo{volume}{78}}, \bibinfo{pages}{7217}
  (\bibinfo{year}{1983}).

\bibitem[{\citenamefont{Eubank and Hall}(1990)}]{virmix:90}
\bibinfo{author}{\bibfnamefont{P.~T.} \bibnamefont{Eubank}} \bibnamefont{and}
  \bibinfo{author}{\bibfnamefont{K.~R.} \bibnamefont{Hall}},
  \bibinfo{journal}{AIChE Journal} \textbf{\bibinfo{volume}{36}},
  \bibinfo{pages}{1661} (\bibinfo{year}{1990}).

\bibitem[{\citenamefont{Bunker}(1979)}]{Bunker-book}
\bibinfo{author}{\bibfnamefont{P.~R.} \bibnamefont{Bunker}},
  \emph{\bibinfo{title}{Molecular Symmetry and Spectroscopy}}
  (\bibinfo{publisher}{Academic Press Inc.}, \bibinfo{year}{1979}).

\bibitem[{\citenamefont{Green}(1975)}]{Green75}
\bibinfo{author}{\bibfnamefont{S.}~\bibnamefont{Green}}, \bibinfo{journal}{J.
  Chem. Phys.} \textbf{\bibinfo{volume}{62}}, \bibinfo{pages}{2271}
  (\bibinfo{year}{1975}).

\bibitem[{\citenamefont{van~der Avoird et~al.}(1994)\citenamefont{van~der
  Avoird, Wormer, , and Moszynski}}]{Avoird:94}
\bibinfo{author}{\bibfnamefont{A.}~\bibnamefont{van~der Avoird}},
  \bibinfo{author}{\bibfnamefont{P.}~\bibnamefont{Wormer}}, , \bibnamefont{and}
  \bibinfo{author}{\bibfnamefont{R.}~\bibnamefont{Moszynski}},
  \bibinfo{journal}{Chem. Rev.} \textbf{\bibinfo{volume}{94}},
  \bibinfo{pages}{1931} (\bibinfo{year}{1994}).

\bibitem[{\citenamefont{Hutson}(1994)}]{Hutson94}
\bibinfo{author}{\bibfnamefont{J.~M.} \bibnamefont{Hutson}},
  \bibinfo{journal}{Comp. Phys. Comm.} \textbf{\bibinfo{volume}{84}},
  \bibinfo{pages}{1} (\bibinfo{year}{1994}).

\bibitem[{\citenamefont{Alexander and DePristo}(1977)}]{Alexander}
\bibinfo{author}{\bibfnamefont{M.~H.} \bibnamefont{Alexander}}
  \bibnamefont{and} \bibinfo{author}{\bibfnamefont{A.~E.}
  \bibnamefont{DePristo}}, \bibinfo{journal}{J. Chem. Phys.}
  \textbf{\bibinfo{volume}{66}}, \bibinfo{pages}{2166} (\bibinfo{year}{1977}).

\bibitem[{\citenamefont{Brown and Carrington}(2003)}]{Carrington-book}
\bibinfo{author}{\bibfnamefont{J.}~\bibnamefont{Brown}} \bibnamefont{and}
  \bibinfo{author}{\bibfnamefont{A.}~\bibnamefont{Carrington}},
  \emph{\bibinfo{title}{Rotational Spectroscopy of Diatomic Molecules}}
  (\bibinfo{publisher}{Cambridge University Press}, \bibinfo{year}{2003}).

\bibitem[{\citenamefont{J.~M.~Hutson}(1993)}]{bound}
\bibinfo{author}{\bibfnamefont{{\sf BOUND}.}~\bibnamefont{J.~M.~Hutson}}
  (\bibinfo{year}{1993}), \bibinfo{note}{computer Code, Version 5, distributed
  by Collaborative Computational Project no. 6 of the Science and Engineering
  Research Council (UK)}.

\bibitem[{\citenamefont{Manolopoulos}(1986)}]{Mano}
\bibinfo{author}{\bibfnamefont{D.~E.} \bibnamefont{Manolopoulos}},
  \bibinfo{journal}{J. Chem. Phys.} \textbf{\bibinfo{volume}{85}},
  \bibinfo{pages}{6425} (\bibinfo{year}{1986}).

\bibitem[{\citenamefont{Johnson}(1973)}]{Johnson}
\bibinfo{author}{\bibfnamefont{B.~R.} \bibnamefont{Johnson}},
  \bibinfo{journal}{J. Comp. Phys.} \textbf{\bibinfo{volume}{13}},
  \bibinfo{pages}{445} (\bibinfo{year}{1973}).

\bibitem[{\citenamefont{Thornley and Hutson}(1994)}]{Thornley}
\bibinfo{author}{\bibfnamefont{A.~E.} \bibnamefont{Thornley}} \bibnamefont{and}
  \bibinfo{author}{\bibfnamefont{J.~M.} \bibnamefont{Hutson}},
  \bibinfo{journal}{J. Chem. Phys.} \textbf{\bibinfo{volume}{101}},
  \bibinfo{pages}{5578} (\bibinfo{year}{1994}).

\bibitem[{\citenamefont{Aquilanti et~al.}(1999)\citenamefont{Aquilanti,
  Ascenzi, Bartolomei, Cappelletti, Cavalli, de~Castro~Vitores, and
  Pirani}}]{Aquilanti:99bis}
\bibinfo{author}{\bibfnamefont{V.}~\bibnamefont{Aquilanti}},
  \bibinfo{author}{\bibfnamefont{D.}~\bibnamefont{Ascenzi}},
  \bibinfo{author}{\bibfnamefont{M.}~\bibnamefont{Bartolomei}},
  \bibinfo{author}{\bibfnamefont{D.}~\bibnamefont{Cappelletti}},
  \bibinfo{author}{\bibfnamefont{S.}~\bibnamefont{Cavalli}},
  \bibinfo{author}{\bibfnamefont{M.}~\bibnamefont{de~Castro~Vitores}},
  \bibnamefont{and} \bibinfo{author}{\bibfnamefont{F.}~\bibnamefont{Pirani}},
  \bibinfo{journal}{J. Am. Chem. Soc.} \textbf{\bibinfo{volume}{121}},
  \bibinfo{pages}{10794} (\bibinfo{year}{1999}).

\bibitem[{\citenamefont{Carmona-Novillo
  et~al.}(2012)\citenamefont{Carmona-Novillo, Bartolomei, Hern{\'a}ndez,
  Campos-Mart{\'\i}nez, and Hern{\'a}ndez-Lamoneda}}]{Carmona-Novillo:12}
\bibinfo{author}{\bibfnamefont{E.}~\bibnamefont{Carmona-Novillo}},
  \bibinfo{author}{\bibfnamefont{M.}~\bibnamefont{Bartolomei}},
  \bibinfo{author}{\bibfnamefont{M.~I.} \bibnamefont{Hern{\'a}ndez}},
  \bibinfo{author}{\bibfnamefont{J.}~\bibnamefont{Campos-Mart{\'\i}nez}},
  \bibnamefont{and}
  \bibinfo{author}{\bibfnamefont{R.}~\bibnamefont{Hern{\'a}ndez-Lamoneda}},
  \bibinfo{journal}{J. Chem. Phys.} \textbf{\bibinfo{volume}{137}},
  \bibinfo{pages}{114304} (\bibinfo{year}{2012}).

\bibitem[{\citenamefont{Aquilanti et~al.}(2002)\citenamefont{Aquilanti,
  Bartolomei, Cappelletti, Carmona-Novillo, and Pirani}}]{n2our}
\bibinfo{author}{\bibfnamefont{V.}~\bibnamefont{Aquilanti}},
  \bibinfo{author}{\bibfnamefont{M.}~\bibnamefont{Bartolomei}},
  \bibinfo{author}{\bibfnamefont{D.}~\bibnamefont{Cappelletti}},
  \bibinfo{author}{\bibfnamefont{E.}~\bibnamefont{Carmona-Novillo}},
  \bibnamefont{and} \bibinfo{author}{\bibfnamefont{F.}~\bibnamefont{Pirani}},
  \bibinfo{journal}{J. Chem. Phys.} \textbf{\bibinfo{volume}{117}},
  \bibinfo{pages}{615} (\bibinfo{year}{2002}).

\end{thebibliography}

\end{document}